\documentclass[journal]{IEEEtran}
\usepackage{cite}
\usepackage[pdftex]{graphicx}
\usepackage{amsmath,amssymb,amsfonts}
\usepackage{balance}
\usepackage{array}
\usepackage{wrapfig}
\usepackage{multirow}
\usepackage{tabu}
\usepackage{csquotes}
\usepackage{xcolor,soul}
\usepackage{url}
\usepackage{float} 
\usepackage{algorithm}
\usepackage{algorithmic}
\usepackage[caption=false, font=footnotesize]{subfig}
\newcommand{\RomanNumeralCaps}[1]{\MakeUppercase{\romannumeral #1}}

\def\BibTeX{{\rm B\kern-.05em{\sc i\kern-.025em b}\kern-.08em
    T\kern-.1667em\lower.7ex\hbox{E}\kern-.125emX}}

\begin{document}

\title{Multi-objective Digital Design Optimisation via Improved Drive Granularity Standard Cells}

\author{Linan~Cao,~\IEEEmembership{Student~Member,~IEEE,}
        Simon~J.~Bale,
        and~Martin~A.~Trefzer,~\IEEEmembership{Senior~Member,~IEEE}
\thanks{The authors are with the Intelligent Systems and Nanoscience Research Group, Department of Electronic Engineering, University of York, York, North Yorkshire, YO10 5DD, United Kingdom. e-mail: (lc1492, martin.trefzer, simon.bale)@york.ac.uk}
}

\maketitle

\begin{abstract}
To tackle the complexity of state-of-the-art electronic systems, silicon foundries continuously shrink the technology nodes and electronic design automation (EDA) vendors offer hierarchical design flows to decompose systems into smaller blocks. However, such a staged design methodology consists of various levels of abstraction, where margins will be accumulated and result in degradation of the overall design quality. This limits the full use of capabilities of both the process technology and EDA tools. In this work, a study of drive granularity of standard cells is performed and an interpolation method is proposed for drive option expansion within original cell libraries. These aim to investigate how industrial synthesis tools deal with the drive strength selection using different granularity sets. In addition, a fully-automated, multi-objective (MO) EDA digital flow is introduced for power, performance, area (PPA) optimisation based on drive strength refinement. This population-based search method better handles the increased difficulty of cell selection when using larger logic libraries, producing better optimised solutions than standard tool flow in this case. The achieved experimental results demonstrate how the improved drive granularity cells overall enhance the quality of designs and how a significant improvement in trading off PPA is achieved by the MOEDA flow.

\end{abstract}

\begin{IEEEkeywords}
Multi-objective Optimisation, Drive Strength, Standard Cells, Digital Flow, EDA.
\end{IEEEkeywords}

\IEEEpeerreviewmaketitle

\section{Introduction}
\IEEEPARstart{S}{tandard} cells, the basic building blocks of digital integrated circuits (ICs), implement basic logic functions. Any large and complex logic function is composed of standard cells from a library providing multiple drive strength options for each cell to meet design specifications. Cells with different drive strengths are realised through different transistor sizes, whereby larger transistors provide increased current drive capabilities and smaller ones consume less area or power. Commercial digital IC design flows commonly use standard cell libraries from foundries, which are predefined for generic design requirements to tape out chips. Well-optimised libraries have therefore become crucial as they determine the overall achievable quality of results (QoR).

The provided drive strength options of a logic cell are limited and therefore of relatively coarse granularity and range. Although limiting and discretising drive options accelerates cell selection to handle modern, large complex designs fast, an optimum scenario would be that EDA tools could select cells of exact drive to meet load requirements thereby avoiding over-design in terms of power and area. Methods like improving drive strength resolution in adjacent most commonly used cells (typically introducing additional smaller sizes) can potentially improve designs particularly for lowering power~\cite{hashimoto2003standard}~\cite{zhang2008standard}~\cite{islam2019drive}.

Technology down-scaling leads to high-density in standard cell layouts that need to accommodate restrictive physical design rules. This incurs long turnaround time with significant human effort in transistor-level placement and routing when creating cell libraries. Automated transistor-sizing tools have been introduced for the provision of fine-tuned drive strength options, mainly focusing on low power design solutions~\cite{hashimoto2001post,vujkovic2002optimized,afonso2009power,rahman2010design}. Furthermore, on-demand transistor sizers~\cite{onodera2001asic}~\cite{northrop2001semi} have been developed for real-time library generation working in the digital EDA flow from logic synthesis to physical design. Such continuously-sized logic cells extend the solution space, but significantly increase the implementation and analysis time~\cite{lavagno2016electronic}. A synthesis-centered design approach, using discrete gate libraries, is still the most commonly used technique for producing new generations of chips in response to the rapid time-to-market process. 

Seeking to achieve richer cell libraries, implementing logic designs using mixed-height (i.e., routing tracks like 9-track and 12-track) or double-row-height standard cells are recently proposed~\cite{dobre2015mixed,chiang2018designing,zhao2019mixed,zhu2020mixed}. Smaller-height cells feature compact area and lower power dissipation, but are weaker in drive strength. Cells with larger heights provide higher cell drive capabilities, but consume more area and power. Mixing different-height cell libraries, available from foundries, is an alternative efficient approach to achieve richer drive options. However, current EDA tools cannot directly handle the mixed-height cell placement legalization so that dedicated place and route tools need to be developed for each case. Interpolating fine-grained drive strength of logic gates based on an existing cell library and inserting them to expand the original granularity can be straightforwardly implemented in standard tools.  This approximates circuit optimisation close to transistor-level, although it might still require custom-design effort, but can ensure the design legalization for fabrication.

However, richer standard cell libraries lead to increasingly difficult logic synthesis when aiming at producing well-optimised technology-mapped netlists. This makes design margins or even errors propagate through the entire flow and ultimately may lead to performance loss due to generic overheads. To resolve these issues, developing a method that can perform optimisation from a more global viewpoint requires more computing time but can provide solutions with considerable improvements.

Population-based optimisation techniques like evolutionary algorithms (EAs) are widely-adapted approaches to perform efficient design space search and provide globally-optimised solutions. Researchers are looking at evolutionary design space optimisation at the system or architecture level~\cite{palesi2002multi}~\cite{ascia2002framework}, intellectual property (IP) blocks~\cite{papamichael2015nautilus}, standard cell library composition~\cite{ricci2007evolutionary}, etc., but limited research is investigating the application of evolutionary optimisation to the full standard digital flow. An automated multi-objective optimisation flow is needed which can compatibly work across different abstraction levels of existing flow to recover overall performance of designs.

Novel contributions made in this work are: 1) A methodology is introduced to interpolate fine-grained drive options into the original granularity of standard cell libraries. This helps to further exploit a technology node, enhancing design quality of digital circuits with better PPA metrics; 2) An MOEDA flow is proposed which applies multi-objective evolutionary algorithms (MOEAs) to a standard digital flow performing multi-objective optimisation and enlarging design solution space based on drive strength mapping. This optimisation approach is used to assist tools to further improve and better trade-off solutions in PPA when synthesising designs with an improved drive granularity library.

The remaining parts of the paper are structured as follows: Section~\ref{section:drive strength} illustrates the design of fine-grained drive strength library. Section~\ref{section:Flow} provides an introduction of MOEDA framework. Experiment setup is described in Section~\ref{section:exp. setup}. Section~\ref{section:results} demonstrates experimental results and Section~\ref{section:conclusion} concludes and outlines future work.

\section{Drive Strength Design of Standard cells}
\label{section:drive strength}
\subsection{Logic Design using Multiple Drive Options}
To have multiple drive strengths for logic functions in a cell library is crucial to achieve timing closure. It is normally indicated using a post-fix after a cell function name, such as X1, X2, X3, etc., in a library. This can provide various drive capabilities to meet different loads when building real circuits at the physical level. Using larger drive strength cells generally consumes more electrical power, die area and pin capacitance, but is able to drive larger loads or increase clock frequency.

Synthesis tools are mapping drive strength from a finite set of discrete drive options for a generic functional gate, and usually need to select the smallest possible one to minimise power consumption and area while trying to meet the timing constraint simultaneously. Hence, a limited number of coarse-grained cell drive strengths will inherently lead to over-design. For example, when a wire delay corresponds to a drive equivalent of X1.5, and choices available are only X1 and X2, then X2 would be selected in order to meet timing requirement~\cite{zhang2008standard}. Improving drive granularity of cells, such as adding some intermediate-sized cells X1.2, X1.5, X2.5, etc., thus could avoid this problem.

Industry-standard cell libraries are designed in a proprietary way. This limits design optimisation to whatever drive strength options are available in a given library to effect design-specific transistor sizing when implementing designs in the digital flow. Pre-designing a significantly richer set of drive strengths for each functional cell can approach a more ideal scenario where the selected drive strength can meet required loads in a more accurate way. However, this would require a huge manual design effort when creating cell libraries, and it is too expensive and time consuming to make libraries even larger than they are already (typically 600-1000 cells).

In prior work,~\cite{islam2019drive} proposes using different drive strength compositions but keeping the original library resolution (i.e., the total number of drive options of each logic gate is fixed) to minimise leakage power consumption especially when circuits operating at relative lower clock frequency or in the sleep mode. This work particularly brings more smaller drive strength which are less than the typical drive strength X1.

However, limited research to date investigates how the synthesis tools deal with the different drive granularity of cells and how this would affect the final results of digital circuits.

\subsection{Improved Drive Granularity Library Design}
\begin{table*}[htbp]
\centering\vspace{-0.3cm}
\caption{Contents of Each Experimental Cell Library}\vspace{-0.2cm}
\begin{tabular}{|c|c|c|}
\hline
Library Name & Functions & Inverters (INV) \\
\hline
MINI\_ORIG & NANDX0 & X0,\hspace{0.72cm} X1,\hspace{0.72cm} X2,\hspace{0.72cm} X3,\hspace{0.72cm} X4,\hspace{0.52cm} X6,\hspace{0.52cm} X8,\hspace{0.65cm} X12,\hspace{0.65cm} X16,\hspace{0.65cm} X20,\hspace{0.65cm} X24\\
\hline
MINI\_FINE & NANDX0 & X0, \textcolor{red}{X0.5}, X1, \textcolor{red}{X1.5}, X2, \textcolor{red}{X2.5}, X3, \textcolor{red}{X3.5}, X4, \textcolor{red}{X5}, X6, \textcolor{red}{X7}, X8, \textcolor{red}{X10}, X12, \textcolor{red}{X14}, X16, \textcolor{red}{X18}, X20, \textcolor{red}{X22}, X24\\
\hline
\end{tabular}\vspace{-0.4cm}
\label{table:expe_libs}
\end{table*}

Our proposed design methodology for improving cell drive resolution is to interpolate custom-designed cells into the original library in the middle of two cells with adjacent drive strengths. Instead of generating a large number of cells with fine-grained drive strength, this method aims to expand drive granularity of cells based on a well-optimised industrial library, and all newly produced cells are aligned with the original library in terms of logic cell drive capabilities.

Here, the TSMC 65nm technology is used, but its pre-designed standard cell library (TCBN65LP) including schematics and full layouts is proprietary and therefore unavailable. Hence, in order to create a representative test case for the 65nm technology used, a reduced library is firstly initialised including 11 inverters (INV) which have the same drive strengths as those in the TSMC TCBN65LP library, and one NAND logic function (NANDX0) with minimum drive strength (transistors are of smallest width). This re-designed cell library, that is modelled to match the original drive granularity of the commercial library, is named ``MINI\_ORIG''.

Subsequently, a set of inverters of more fine-grained drive strengths are interpolated into ``MINI\_ORIG'' to form another library named ``MINI\_FINE''. Both custom-designed libraries and their drive granularity are summarised in Table~\ref{table:expe_libs}. To focus the investigation on the drive strength selection and simplify the problem, we only consider drive strength expansion of inverters in this case.

To define the drive strength of a gate needs to be based on its performance evaluation (i.e., the speed to drive a load capacitance). For example, if the X1 can drive a unit load capacitance $\mathrm{C_{unit\_load}}$ in a period time $\mathrm{T_{unit\_load}}$ (i.e., circuit delay), the X2 needs to be designed through iterative transistor-sizing until it can drive double the unit load capacitance $2\times\mathrm{C_{unit\_load}}$ taking a near-exact same time $\mathrm{T_{unit\_load}}$. So the definition of drive strength:

\begin{equation}
\begin{aligned}
\displaystyle & \mathrm{X} = \dfrac{\mathrm{C_{X\_load}}}{\mathrm{C_{unit\_load}}} \quad \textrm{s.t.} \quad \mathrm{T_{X}} = \mathrm{T_{unit\_load}}
\end{aligned}
\end{equation}

In addition, the transistor size of drive strength X1.5 in ``MINI\_FINE'' library is defined when it can drive the $1.5\times\mathrm{C_{unit\_load}}$ in the same time $\mathrm{T_{unit\_load}}$. The following drive strengths in both ``MINI\_ORIG'' and ``MINI\_FINE'' like X2.5, X3, X3.5, X4, etc., are all created using the same approach.
 
The inverter drive strength X1 is defined by $\mathrm{PMOS}_\mathrm{size}=\mathrm{230nm}$ and $\mathrm{NMOS}_\mathrm{size}=\mathrm{165nm}$, so the $\mathrm{P/N}$ ratio adapted in this work is $1.39$ for all cells. The X0 cell is defined by $\mathrm{NMOS}_\mathrm{min}$ and $\mathrm{PMOS}_\mathrm{size}=\mathrm{1.39}\times\mathrm{NMOS}_\mathrm{min}$ according to the minimum design rules from the technology. The X0.5 inverter is then interpolated in the middle between X0 and X1 through transistor-sizing until it can drive a load capacitance value in the middle between X0 and X1's in $\mathrm{T_{unit\_load}}$. All created cells keep the same transistor length $\mathrm{60nm}$.

\begin{figure}[htbp]
    \centering\vspace{-0.2cm}
    \includegraphics[width=0.55\linewidth]{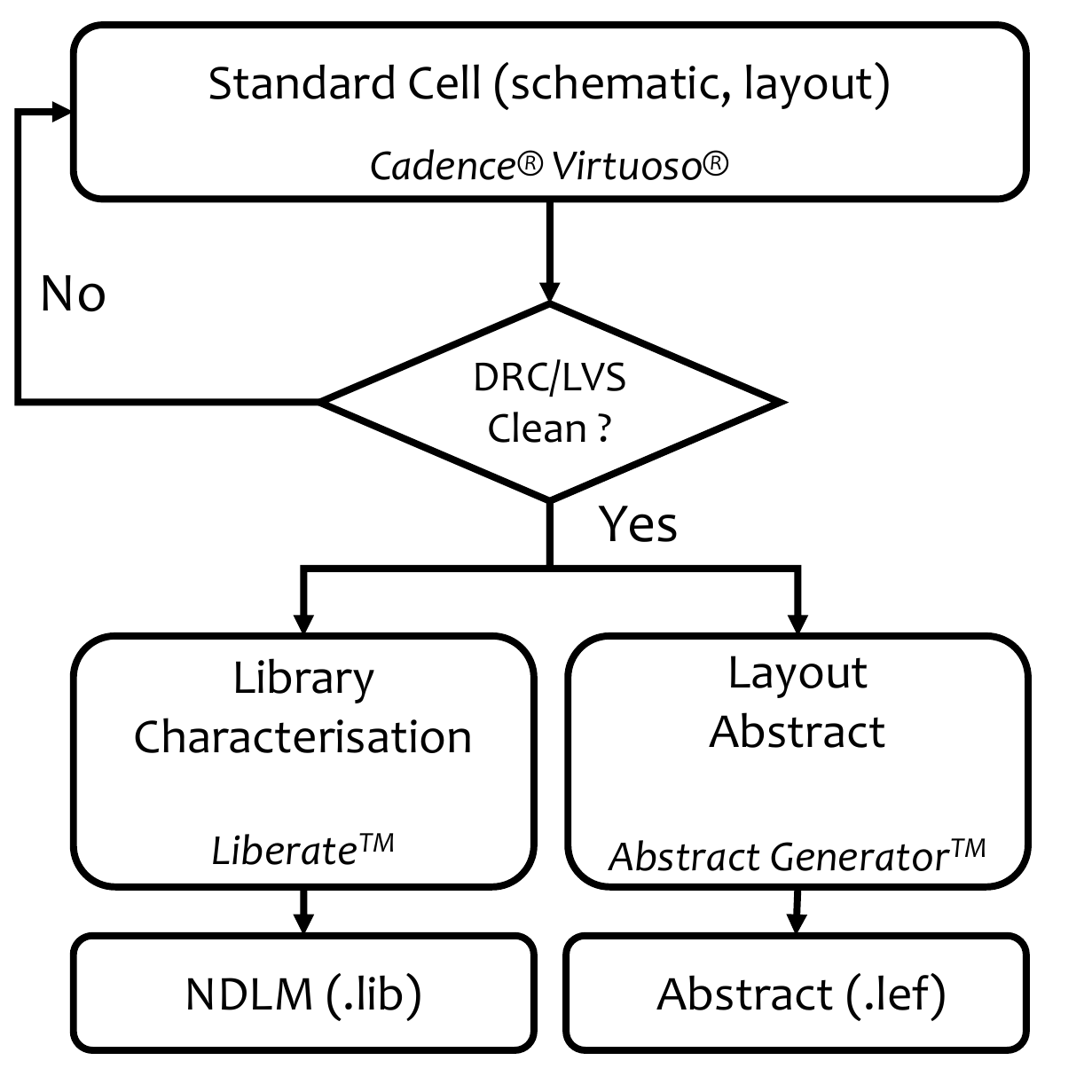}\vspace{-0.2cm}
    \caption{Standard cell design flow including library characterisation and layout abstract.}
    \label{fig:cell_flow}
\end{figure}

In this work, both custom-designed ``MINI\_ORIG'' and ``MINI\_FINE'' libraries are implemented including schematics and layouts using Cadence\textsuperscript{\textregistered} Virtuoso\textsuperscript{\textregistered}. All library cells are designed in body tapped structure. The cell layouts are characterised respectively into timing and power models (Liberty file) and physical abstractions (LEF file, top layer view of layouts) using using Cadence\textsuperscript{\textregistered} Liberate\textsuperscript{TM} and Abstract Generator\textsuperscript{TM} tools. The standard cell design flow is illustrated in Fig.~\ref{fig:cell_flow}.

The Non-Linear Delay Model (NLDM) is a look-up-table-based model containing timing and power information of each gate. The two input indexes given are \textit{input slew} and \textit{load capacitance}. The output index is the circuit's delay or power under different compositions of the input slew and the load capacitance to separately produce delay and power look-up tables. Both delay and power are evaluated through a series of SPICE-based simulations run by the Liberate characterisation tool.

The Liberty (.lib) file contains two main parts: the first one contains the technology library including all environment descriptions like operating conditions, wire load mode, etc., and the second one contains the cell descriptions obtained by running the library characterisation tool. The technology library, in this case, is using the typical corner (PVT: TT, 1.2V, 25$^{\circ}$C) of TSMC65nm technology, which is the same as the TCBN65LP library uses, and the environment descriptions are kept the same as well. In addition, a 7x7 look-up-table NLDM is used for cell descriptions and characterisation input index values are inferred from the original TCBN65LP library.

The input slew in this case is a fixed range where the input signal transition ranges from a close-ideal step to a larger slew time. The same input slew set is used for all cells. Furthermore, each designed drive strength has a specific capacitive load set ranging from a small to large capacitance value. The inverters in ``MINI\_ORIG'' library use the corresponding load capacitance indexes from the TCBN65LP library, but the load capacitance index for each fine-grained inverter is found through calculating the middle (or average) value of two adjacent cells' load capacitance indexes.

The characterised information of each gate includes both timing and power consumption tables. The timing tables of each gate include cell delay (i.e., measured from 50\% to 50\%) and transition time (i.e., measured from 30\% to 70\% in this case). Both the cell delay and transition time are specified during the characterisation. Hence, four tables per input pin of a logic gate are generated, including cell rise, cell fall, rise transition and fall transition. The power consumption information in the NLDM includes two parts: internal power and leakage power. The internal power, or called short-circuit power, is the power dissipated by an instantaneous short-circuit current flowing between the supply voltage and the ground at the time the gate switches state. The power dissipation table describes each cell's internal power consumption as the combination of energy consumed by output and input pin transitions with respect to a given clock frequency. The values provided represent the amount of energy consumed (in $\mathrm{uW/MHz}$ or $\mathrm{pJ}$) within the cell when the corresponding output pin state changes. Input pin energy consumption is included to increase accuracy of estimated power consumption, where the consumed energy value is measured for each input pin toggle while output pin state remains unchanged.  In order to obtain the internal power consumption, the consumed energy needs to be considered with a clock frequency applied in the EDA tool's power analysis. The average/min/max leakage power values are provided in nanowatts ($\mathrm{nW}$) for immediate use.

\subsection{The Performance of the Proposed Libraries}
\begin{table}[htbp]
\centering
\caption{Library Cell Information}\vspace{-0.2cm}
\begin{tabular}{|l|c|c|c|c|c|}
    \multicolumn{6}{r}{PVT: TT, 1.2V, 25$^{\circ}$C}\\\hline
    \multicolumn{1}{|c|}{Cell} & Transistor & Cell       &\multicolumn{3}{c|}{Leakage Power [$nW$]} \\\cline{4-6}
    \multicolumn{1}{|c|}{Name} & Count      & Width [$um$] & Min. & Ave. & Max. \\\hline
    INVX0    & 2  & 0.6 & 0.0106 & 0.0108 & 0.0111 \\\hline
    INVX0.5  & 2  & 0.6 & 0.0114 & 0.0117 & 0.0120 \\\hline
    INVX1    & 2  & 0.6 & 0.0112 & 0.0133 & 0.0153 \\\hline
    INVX1.5  & 2  & 0.7 & 0.0184 & 0.0225 & 0.0265 \\\hline
    INVX2    & 4  & 0.8 & 0.0282 & 0.0297 & 0.0311 \\\hline
    INVX2.5  & 4  & 1.0 & 0.0388 & 0.0406 & 0.0423 \\\hline
    INVX3    & 6  & 1.2 & 0.0511 & 0.0516 & 0.0522 \\\hline
    INVX3.5  & 6  & 1.3 & 0.0627 & 0.0637 & 0.0646 \\\hline
    INVX4    & 8  & 1.4 & 0.0704 & 0.0731 & 0.0759 \\\hline
    INVX5    & 10 & 1.6 & 0.0894 & 0.0948 & 0.1002 \\\hline
    INVX6    & 12 & 1.8 & 0.1070 & 0.1167 & 0.1264 \\\hline
    INVX7    & 14 & 2.2 & 0.1301 & 0.1441 & 0.1581 \\\hline
    INVX8    & 16 & 2.4 & 0.1497 & 0.2575 & 0.1853 \\\hline
    INVX10   & 20 & 3.0 & 0.1931 & 0.2191 & 0.2451 \\\hline
    INVX12   & 24 & 3.4 & 0.2325 & 0.2673 & 0.3022 \\\hline
    INVX14   & 28 & 4.0 & 0.2766 & 0.3201 & 0.3636 \\\hline
    INVX16   & 32 & 4.4 & 0.3165 & 0.3686 & 0.4207 \\\hline
    INVX18   & 36 & 5.0 & 0.3605 & 0.4223 & 0.4841 \\\hline
    INVX20   & 40 & 5.6 & 0.4044 & 0.4760 & 0.5475 \\\hline
    INVX22   & 44 & 6.2 & 0.4482 & 0.5295 & 0.6109 \\\hline
    INVX24   & 48 & 6.6 & 0.4885 & 0.5790 & 0.6695 \\\hline
    NANDX0   & 4  & 0.8 & 0.0028 & 0.0152 & 0.0303 \\\hline
\end{tabular}\vspace{-0.5cm}
\label{table:cell_info}
\end{table}

To verify whether the proposed libraries are designed in an appropriate way, and particularly the fine-grained drive inverters are properly interpolating into the original granularity, the delay and power of each gate are analysed to determine the relationship between two adjacent drive strengths and the overview of all cells. 

Table~\ref{table:cell_info} shows the transistor count, cell width and leakage power (i.e., specifically contains average/min/max values) of each cell for both ``MINI\_ORIG'' and ``MINI\_FINE'' libraries. Based on the characterisation results, the leakage power increases linearly when the transistor width increases with each drive strength. All cells are created at the same height $\mathrm{1.8um}$, so the cell area also increases as the cell width increases from small to large drive strength, as transistors of increasing size need to be accommodated. Drive strengths X0 and X0.5 have the same width as X1 due to constraints of the physical design rules. Hence, in this case, swapping the three smallest cells provides a gain in power reduction and not in area.

The characterised cell propagation delays confirm that fine-grained drive inverters are interpolating into the original drive granularity in an appropriate way. All inverters can drive the specified sets of loads with approximately same speed. Exceptions are X0 and X0.5 which, due to the minimum physical design constraints, cannot be down-sized further. Regardless of that, the X0.5 inverter is properly interpolated between X0 and X1. In terms of power consumption, all fine-grained drive inverters are also positioned in the middle of adjacent original granularity inverters.

Following on from the analysis and discussion of both custom-designed libraries, these will be firstly loaded into the standard digital flow in order to investigate how the EDA tools trade-off design solutions when using a rich (finer-grained) drive strength library compared to the original, coarser-grained one. The MOEDA flow will then optimise drive strength selection based on the tool-optimised gate-level netlists in order to search for better solutions in PPA.

\section{MODEA Optimisation Framework}
\label{section:Flow}
\subsection{Standard Digital Flow}
Modern digital IC design flow, a solid and mature process, consists of various steps including register-transfer-level (RTL) design, logic synthesis, physical implementation (Place and Route) and sign-off (pre-fabrication testing and verification)~\cite{lavagno2016electronic}~\cite{sengupta2016design}. Although the commercial design kit is indeed powerful enough to tackle complex systems, there is still a need for significant human effort involved in the design process. If design violations cannot be resolved at the physical design stage through engineer-change-order (ECO) optimisation, engineers turn back to tuning synthesis, or even design adjustments in components or constraints at the system level, to achieve design closure. This is an extremely time-consuming cycle, with the overall design optimisation challenge to find possible optimal trade-off solutions in regard to multiple design requirements using appropriate library cells while reducing turnaround time~\cite{kahng2011vlsi}.

\subsection{Discrete Gate-sizing}
Selecting the appropriate size for a logic gate to implement circuits down to physical level from discrete libraries is the crucial step to achieve an efficient design and timing closure. The optimisation goal is to minimise power consumption while meeting all timing constraints.

Such constrained optimisation problems have been researched for decades and many approaches to solve them have been proposed. Lagrangian Relaxation (LR) formulation, a mathematical theory, has been established in gate sizing problem with low runtime~\cite{flach2014effective}~\cite{sharma2019lagrangian}. The problem is then simplified by weighted factors (Lagrangian multipliers) that moves the constraints into the primal objective function. It is common to simplify circuit models and solve an abstract, so that a continuous version of the gate-sizing can facilitate convex optimisation problems~\cite{hu2012sensitivity}. This approach does not quite generalise in practice, because device physics often imply non-convex delay functions, causing non-convexity in SPICE results and nonlinear delay model (NLDM) tables~\cite{kahng2013high}.

Practical approaches like~\cite{hu2012sensitivity} used sensitivity guided greedy metaheuristic to reduce timing violations and then minimise the leakage power. In earlier works, typical heuristic techniques like genetic algorithms were applied to solving gate sizing problems~\cite{wang1995performance}~\cite{benkhider2000parallel} based on weighted sum functions. 

Gate sizing problem is multi-objective in nature. Most introduced methods are scalarising based to decompose the optimisation complexity, that combines objectives in one function (e.g., typical weighted sum method). This makes searching highly-efficient but limits achieving feasible Pareto-optimal solutions~\cite{wang2016localized}.

More recently, gate-sizing-based soft error optimisation using MOEAs is proposed~\cite{sheng2009soft} but its multi-objectives are soft error rate, critical path delay and area. In this work we apply MOEAs in a state-of-the-art digital EDA flow to perform drive-strength-mapping-based design space exploration offering a wide range of Pareto-optimised solutions. The optimisation is enhancing already well-optimised solutions generated by tools.

\subsection{Evolutionary Algorithms}
Evolutionary algorithms (EAs) are a class of population-based metaheuristic optimisation algorithms inspired by biological mechanisms like evolution, reproduction, genetics and natural selection. An EA normally starts with an initial~\textit{population}, consisting of $N$ \textit{individuals} (candidate solutions), which is allowed to breed with each evolutionary cycle (\textit{generation}). The initial population can be either randomly initialised or seeded with a set of specific configurations. During each generation, individuals can be modified based on their \textit{chromosomes} through genetic operations such as \textit{mutation} or \textit{crossover} (recombination with each other). All individuals are evaluated using a~\textit{fitness} function and ranked according to their fitness score at the end of each generation. Only the fittest individuals survive the selection process forming the subsequent generation. Termination of the evolution process is triggered when specific criteria are met, e.g., sufficient quality of solution or maximum number of generations.

Implementation of EAs requires:

\textit{(1) Definition of representation.} This is the data structure that the EA manipulates. It represents individuals as a set of genes, the~\textit{chromosome}, comprising all variables and parameters necessary to describe it.

\textit{(2) Implementation of genetic operators.} Mutation and crossover are commonly applied during evolution process. Mutation modifies genes of individuals, and crossover combines subsets of genes of multiple individuals to produce new ones.

\textit{(3) Definition of a fitness function.} This function is used to calculate a fitness score for each individual based on its performance in design objectives. Fitness scores are used during the ranking and selection process to determine which individuals survive to form the population for the next generation.

\subsection{Multi-objective (MO) EDA Digital Flow}

Technology cell mapping is a sub-step in logic synthesis to specify which drive strength would be selected and mapped to each generic functional gate. Gate-level netlists are then generated after the completion of synthesis, where our proposed MOEDA flow focuses on. 

\begin{figure}[htbp]
    \centering
    \includegraphics[width=0.6\linewidth]{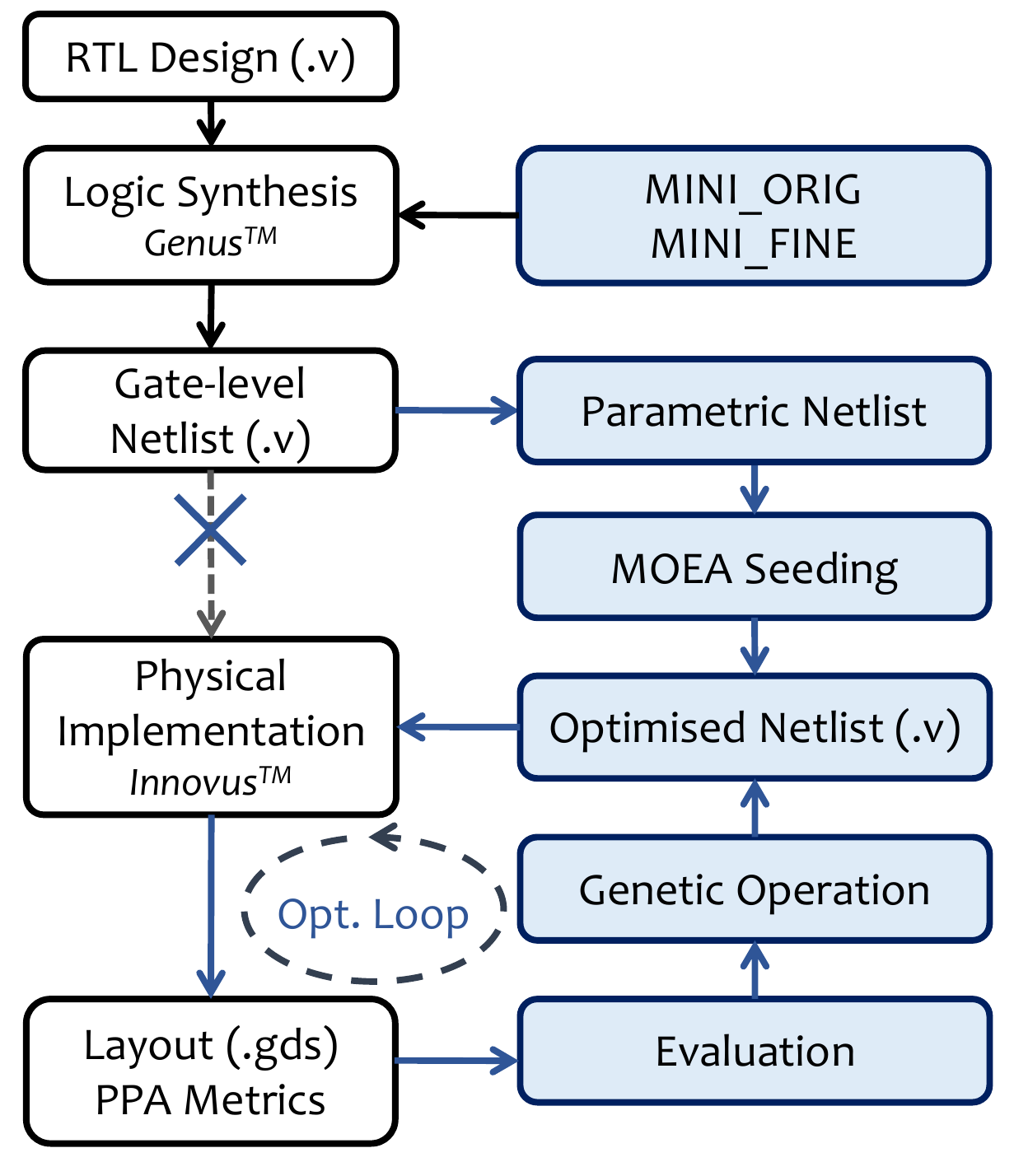}\vspace{-0.2cm}
    \caption{MOEDA digital flow. The flowchart on the left side illustrates a standard digital flow. The MO evolutionary optimisation engine is shown on the right. The blue cross indicates the position where we break the standard flow. The custom-designed cell libraries are used in this flow instead of using the foundry libraries.}\vspace{-0.1cm}
    \label{fig:MOO_flow}
\end{figure}

Fig.~\ref{fig:MOO_flow} presents the proposed flow. A multi-objective evolutionary optimisation loop is tapped between logic synthesis and physical implementation. The MOEDA optimiser automatically performs fine-tuning on drive strength selection of logic gates (i.e., inverters in this case) based on produced synthesised netlists. Logic gates are then placed and routed on physical layouts where the evaluation is performed. The proposed flow specifically involves few steps:

First step is to produce the~\textit{Parametric Netlist} of a synthesised gate-level design. This parameterisation process specifically encodes the drive strengths of inverters into a set of representations, chromosome $\mathbf{g}$, allowing the MOEA to modify them. In this case, a list of integer parameters define each inverter's drive strength to form a parametric netlist.

Second step is to install an initial population, called~\textit{MOEA seeding}. In this work, a solution optimised from the synthesis tool is chosen as the seed. So, seeding a population with a specific solution means all individuals are directly assigned with that solution. This is also the starting point of the MO optimisation process.

After initialising a population, the~\textit{Genetic Operation} only performs the modifications on inverters' drive strengths. Only mutation operator is implemented in this work based on a variation probability $\rho$ to indicate how many inverters out of all will be modified. A mutation results in a modified new netlist ready for physical layout implementation. The evolutionary optimisation loop is then iteratively (over a number of generations) updating netlists for increasingly-optimised solutions.

\textit{Evaluation} is performed using the multi-objective fitness function from equation (2). Metrics (worst case delay $D_{wc}$, total power consumption $P_{total}$ and all gate area $A_{gate}$) are calculated by the EDA tool based on the circuit layout instance.
\begin{equation}
\begin{aligned}
\displaystyle f(\mathbf{g}) = \min \quad & [D_{wc}(\mathbf{g}), \quad P_{total}(\mathbf{g}), \quad A_{gate}(\mathbf{g})]\\
\textrm{s.t.} \quad &  \mathbf{g}=(g_1,...,g_i), \quad \forall g_{i} \in \mathbf{G}
\end{aligned}
\end{equation}

\begin{figure}[htbp]
    \centering\vspace{-0.1cm}
    \includegraphics[width=\linewidth]{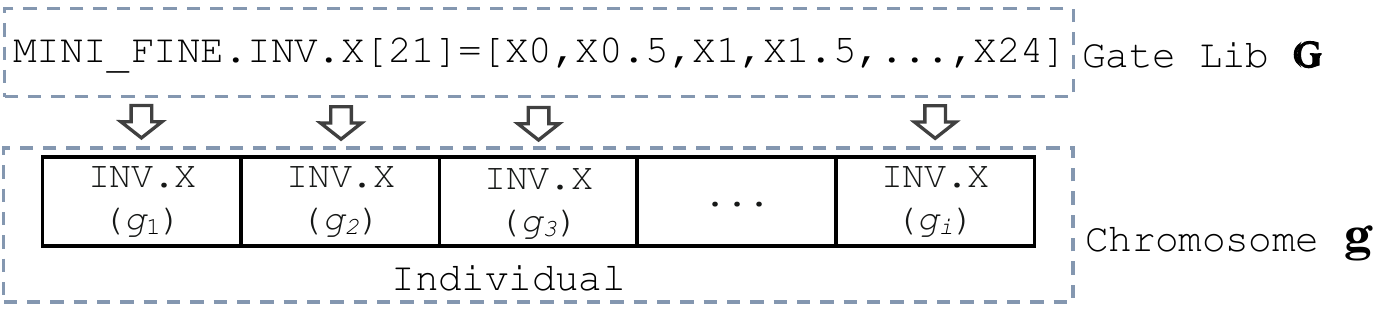}\vspace{-0.3cm}
    \caption{A chromosome example of an individual (i.e., layout instance in this case). All individuals comprise a population.}
    \label{fig:chrom_exam}\vspace{-0.1cm}
\end{figure}

The chromosome vector $\mathbf{g}$ represents the input variables to the fitness functions, which in this case are drive strengths of inverters ($g_{i}$) available from the ``MINI\_FINE'' library ($\mathbf{G}$). Fig.\mbox{~\ref{fig:chrom_exam}} demonstrates a chromosome example of an individual where the $\mathbf{g}=(g_1,...,g_i)$ represents all inverters of it. Each single $g$ (INV.X) shows the drive strength of an inverter. When mutation is triggered, the inverters to be mutated are randomly selected by the MOEA based on the given mutation rate $\rho$. For each selected inverter, the algorithm will randomly choose a new one from $\mathbf{G}$ (including all drive options) to replace the previous one.

In this work, NSGA-\RomanNumeralCaps{2}, one of most popular MOEAs~\cite{deb2002fast}, has been adapted as the searching tool. The~\textit{Non-Dominated-Sorting} and~\textit{Diversity Preservation} are introduced to ensure convergence while achieving a uniform spread of Pareto-optimised solutions.

\textit{Non-Dominated-Sorting}. This is a ranking scheme to evaluate individuals according to their domination level. If an individual $p$ performs better than another $q$ in at least one objective while no degradation in any other objectives, $p$ is said to dominate $q$. In non-dominated-sorting, each individual (e.g., $p$) has a domination count, the number of solutions that dominate $p$. The individuals are grouped based on their domination count into multiple fronts $\mathbf{F} = (\mathbf{F}_{1},...,\mathbf{F}_{i})$. The non-dominated individuals which have the lowest domination counts (i.e., zero) form the first front $\mathbf{F}_{1}$. The individuals which have the second lowest domination counts form the second front $\mathbf{F}_{2}$ and this will continue to the third and following fronts until all individuals are assigned.

\begin{algorithm}[ht!]
\caption{Evolutionary Optimisation in the MOEDA Flow}
\textbf{Procedure:} NSGA-\RomanNumeralCaps{2}($N$, $M$, $f(\mathbf{g})$). $\triangleright$ $N$ individuals evolved $M$ generations to solve $f(\mathbf{g})$.
\begin{algorithmic}[1]
\STATE Initialize parent population $\mathbf{P}_{t}$ in size $N$ $\triangleright$ Seed with a synthesis-optimised solution generated by the tool.
\STATE Offspring population $\mathbf{Q}_{t}$ $\gets$ Mutation($\mathbf{P}_{t}$)
\FOR{$t \gets 1$ to $M$}
\FOR{each population $\mathbf{R}_{t} \gets \mathbf{P}_{t} \cup \mathbf{Q}_{t}$ in size $2N$}
\STATE Fitness evaluation $\gets$  $f(\mathbf{\mathbf{R}_{t}})$ $\triangleright$ Call fitness function $f(\mathbf{g})$ for each individual evaluation.
\STATE $\mathbf{F}$ $\gets$ Non-Dominated-Sorting($\mathbf{R}_{t}$)
\STATE $\mathbf{P}_{t+1} \gets$ \O
\STATE $i \gets 1$
\WHILE{$|\mathbf{P}_{t+1}| + |\mathbf{F}_{i}| \leq N$}
\STATE Crowding-Distance-Assignment($\mathbf{F}_{i}$)
\STATE $\mathbf{P}_{t+1}$ $\gets$ $\mathbf{P}_{t+1} \cup \mathbf{F}_{i}$
\STATE $i \gets i + 1$
\ENDWHILE
\STATE $\mathbf{F}_{i} \gets$ Descend-Sort($\mathbf{F}_{i}$)
\STATE $\mathbf{P}_{t+1} \gets \mathbf{P}_{t+1} \cup \mathbf{F}_{i}[1:(N-|\mathbf{P}_{t+1}|)]$ $\triangleright$ Less crowned individuals from the 1st to the $(N-|\mathbf{P}_{t+1}|)$th of $\mathbf{F}_i$ to fill $\mathbf{P}_{t+1}$.
\STATE $\mathbf{Q}_{t+1} \gets$ Mutation($\mathbf{P}_{t+1}$)
\ENDFOR
\ENDFOR
\end{algorithmic}
\end{algorithm}

\textit{Diversity Preservation}. This strategy estimates the solution density in the vicinity of each individual based on the Euclidean distance to their nearest neighbours. This needs to firstly calculate the distance of each individual to others and make \textit{Crowding-Distance-Assignment} to each individual, then \textit{Descend-Sort} the $\mathbf{F}$ according to the distance values. If two individuals belong to the same non-dominated front, the one that resides in the less crowded region is preferred.

Algorithm 1 illustrates the overall optimisation process aging with a population in NSGA-\RomanNumeralCaps{2}. The MOEDA flow is continuously producing different circuit instances and keeping elitist ones generation-by-generation, then ultimately achieve a set of wide spread of optimised trade-offs in all objectives.

\section{Experiment Setup}
\label{section:exp. setup}
We implement the proposed algorithm in C++ and all experiments are running on a 2.2GHz Xeon E5-2650 CPU. Test circuits (RTL designs), from ISCAS85 benchmark suite~\cite{iscas}, are synthesised into gate-level netlists using Cadence\textsuperscript{\textregistered} Genus\textsuperscript{TM} (v17.11). Cadence\textsuperscript{\textregistered} Innovus\textsuperscript{TM} (v17.11) tool completes the physical implementation, producing layout instances.

\subsection{Tool Environment Setup}
In order to take full advantage of the built-in optimisations of current EDA tools, it is necessary to push the limits of what tools can achieve with end user-accessible design options. Thus, the~\textit{synthesis-compile-effort} is set to high and~\textit{ultra-optimisation} is enabled in all experiments presented. In addition, in the timing constraint setup in the Genus\textsuperscript{TM} tool, an ideal clock is created running at $\mathrm{250MHz}$ for all inputs and outputs, which means all circuit paths are clocked with two virtual flip-flops from the beginning to the end of each path. The timing constraint is the required time that designs need to meet so the worst path arrival time should be less than the required time (i.e., $\mathrm{4ns}$ in this case).

\begin{figure}[htbp]
    \centering\vspace{-0.2cm}
    \includegraphics[width=\linewidth]{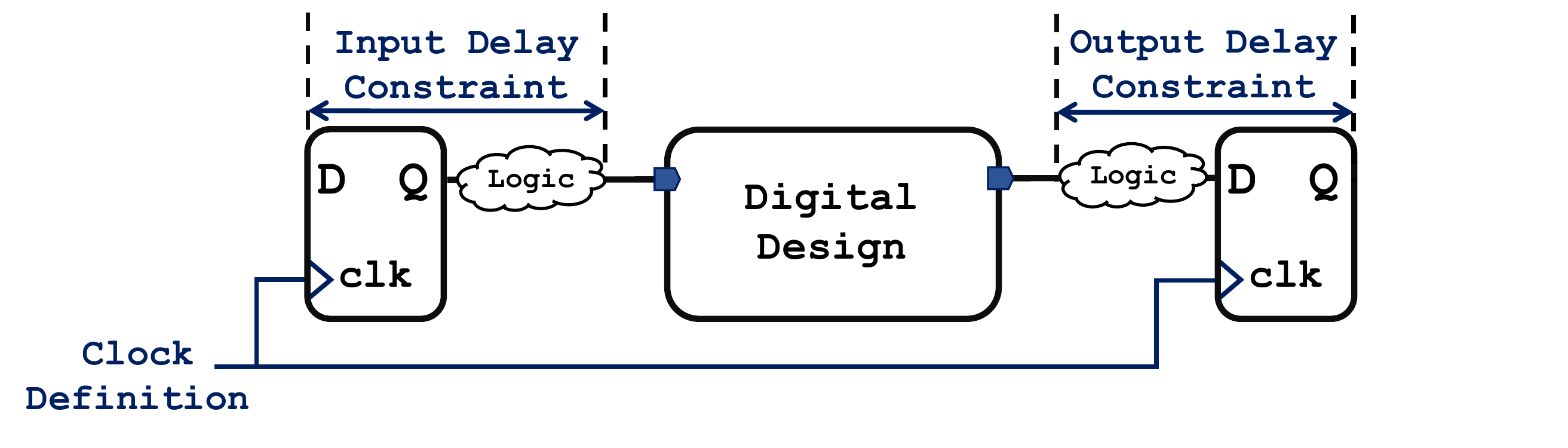}\vspace{-0.3cm}
    \caption{Conceptual testbench to define timing constraints in EDA tools. Virtual logic parts and flip-flops allow the user to specify delays and clocks in the testbench. The design under test is \textit{Digital Design} in the middle.}\vspace{-0.1cm}
    \label{fig:timing_constraints}
\end{figure}

To tighten the timing constraint, the output delay is gradually increased as shown in Fig.~\ref{fig:timing_constraints}. In this work the output delay constraint is increased in increments of 0.05ns, starting from 0ns. The \textit{Required Timing} is obtained by \textit{Clock Period Time - Output Delay Constraint}. Input delay constraints are not applied in this work. The test cases used in this work are combinational circuits, thus, the clock is ideal without any uncertainties or transition delays.

The environment electrical constraint is applied by setting drive strength X1 and X4 output loads. The specific values chosen correspond to the respective inverter X1 and X4's input pin capacitance from ``MINI\_FINE'' library.

\begin{table}[ht!]
\begin{center}\vspace{-0.2cm}
\caption{Design Constraint and Tool Settings in Digital Flow}\vspace{-0.1cm}
\begin{tabular}{|c|c|}
  \hline
  Synthesis Setup & Place \& Route Setup\\ 
  \hline
                                                    & \\
  \multirow{1}{*}{syn\_generic\_effort = high}      & aspect ratio = 1.0 \\
  \multirow{1}{*}{iopt\_ultra\_optimisation = true} & core utilisation = 0.7 \\
  \multirow{3}{*}{Design Constraint}                & noPrePlaceOpt = true \\\cline{1-1}
  \multirow{4}{*}{set\_load = X1/X4}                & timing-driven placement = true \\\cline{1-1}
  \multirow{4}{*}{create\_clock = 250MHz}           & timing-driven routing = true \\
                                                    & \\
                                                    & \\\hline
\end{tabular}\vspace{-0.4cm}
\label{table:tool_setting}
\end{center}
\end{table}

All setup parameters of design constraint, synthesis, place and route are summarised in Table~\ref{table:tool_setting}. The die shape ratio is set to 1.0 and the core utilisation is set to 70\%. The pre-place optimisation, \textit{PrePlaceOpt}, that is to delete buffers or inverters on the gate-level netlists before the placement is disabled in this case. Because we are investigating how the tools select cells, it is worthwhile to keep netlists consistent during both synthesis and physical design steps. Both timing-driven placement and routing are enabled to make designs achieve the best timing that the tools can achieve automatically.

\begin{figure*}[t] 
  \centering\vspace{-0.5cm}
     \subfloat{\includegraphics[width=0.325\linewidth]{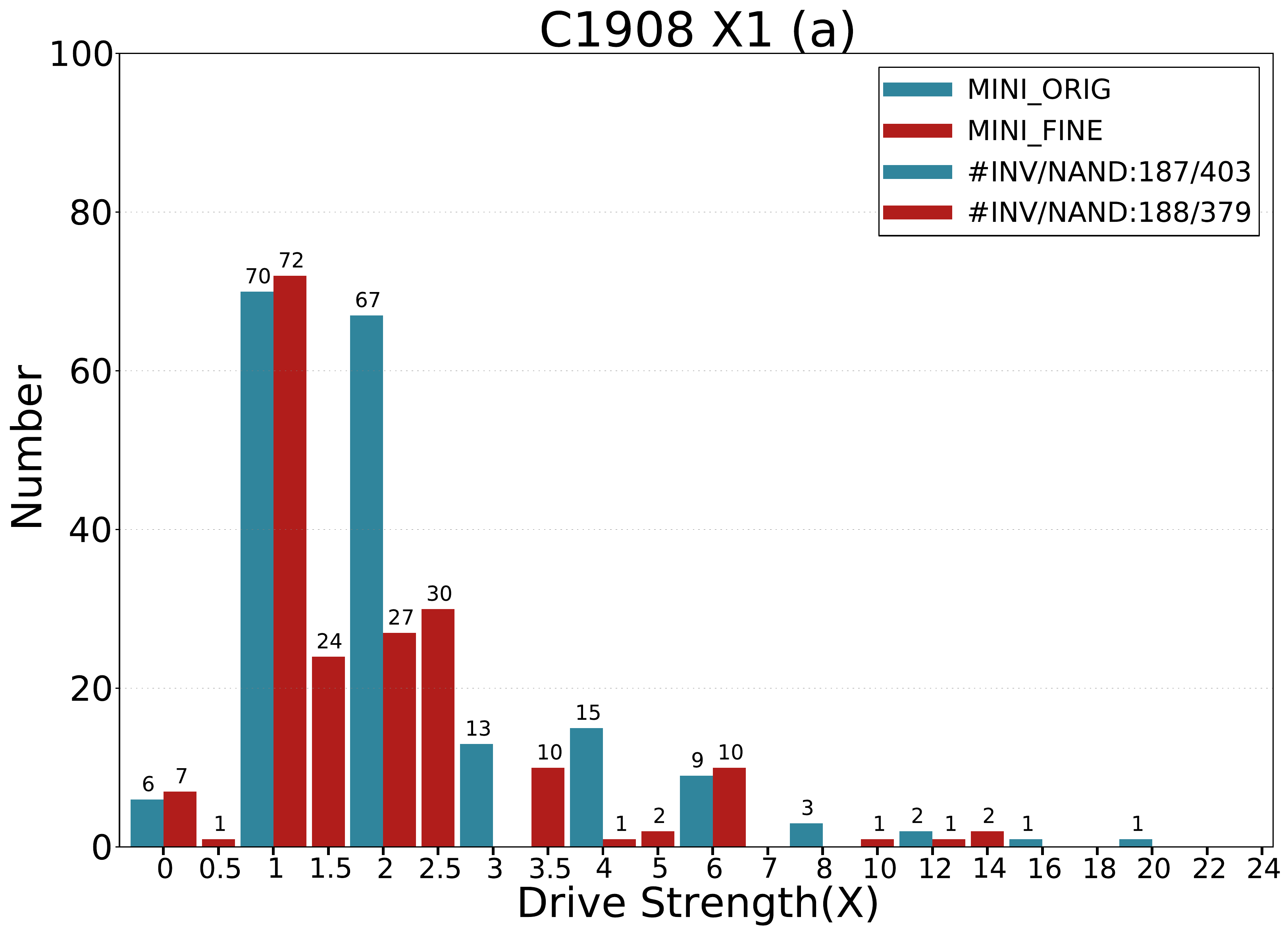}}
     \subfloat{\includegraphics[width=0.325\linewidth]{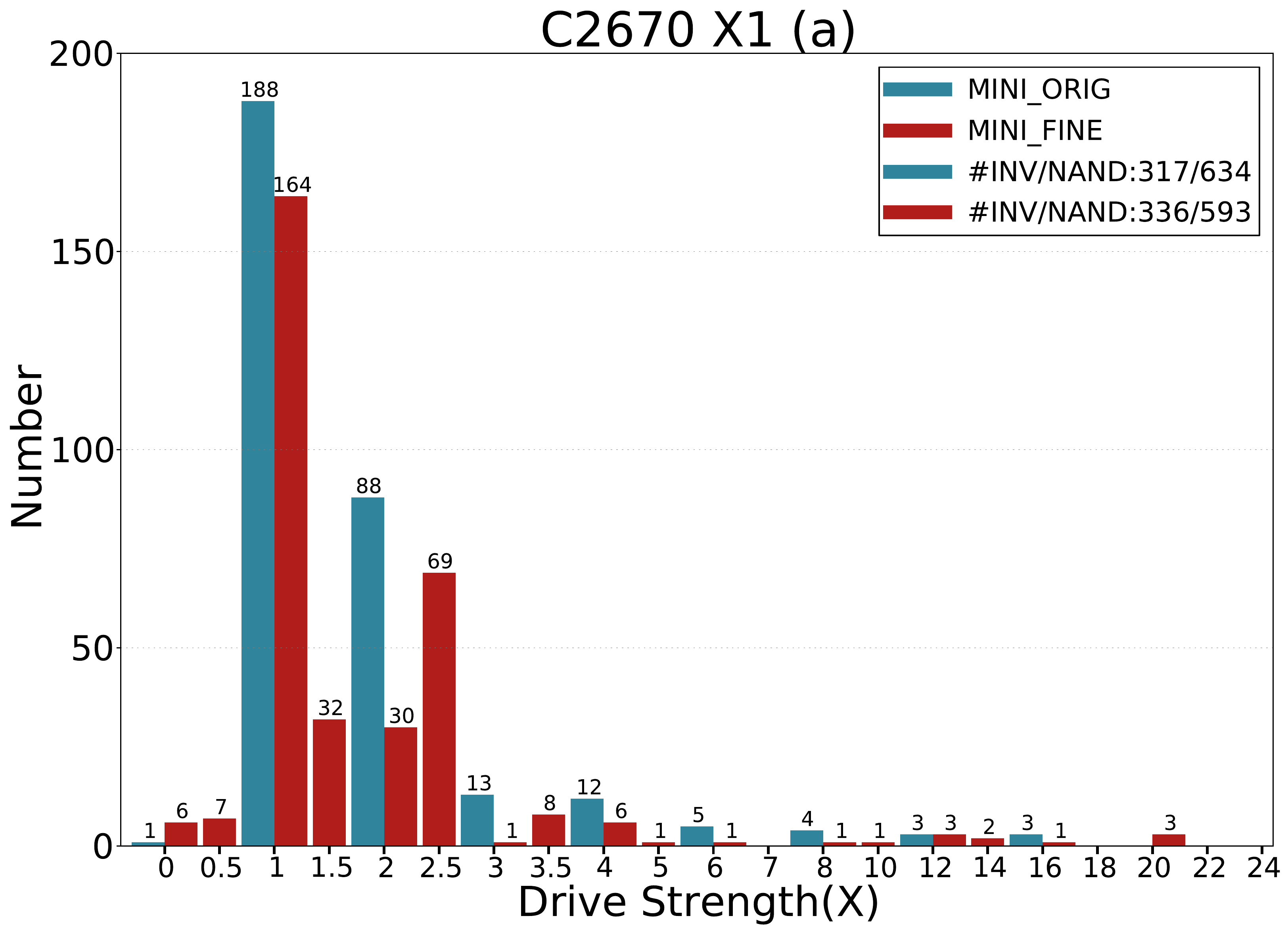}}
     \subfloat{\includegraphics[width=0.325\linewidth]{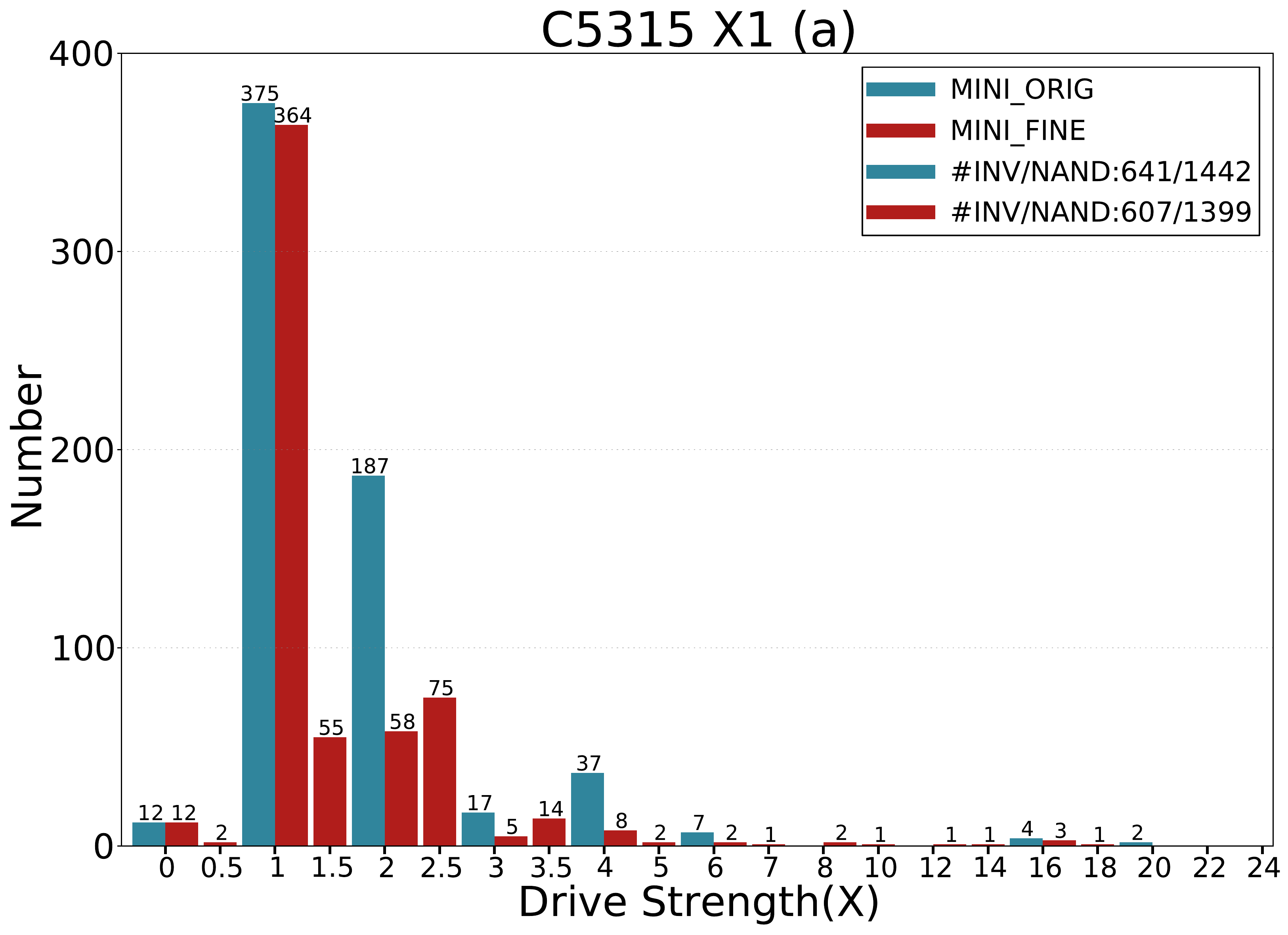}}\\\vspace{-0.3cm}
     \subfloat{\includegraphics[width=0.325\linewidth]{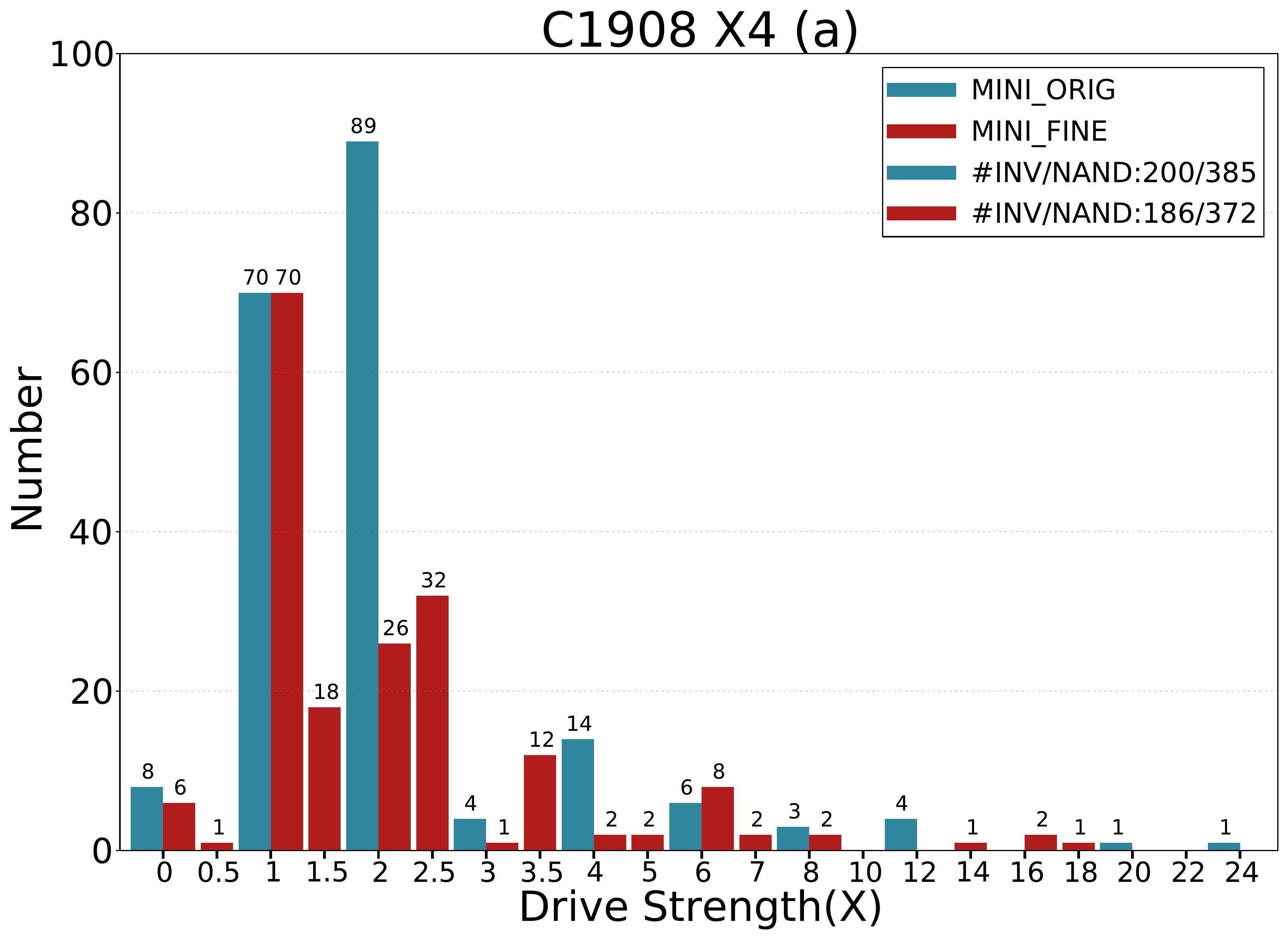}}
     \subfloat{\includegraphics[width=0.325\linewidth]{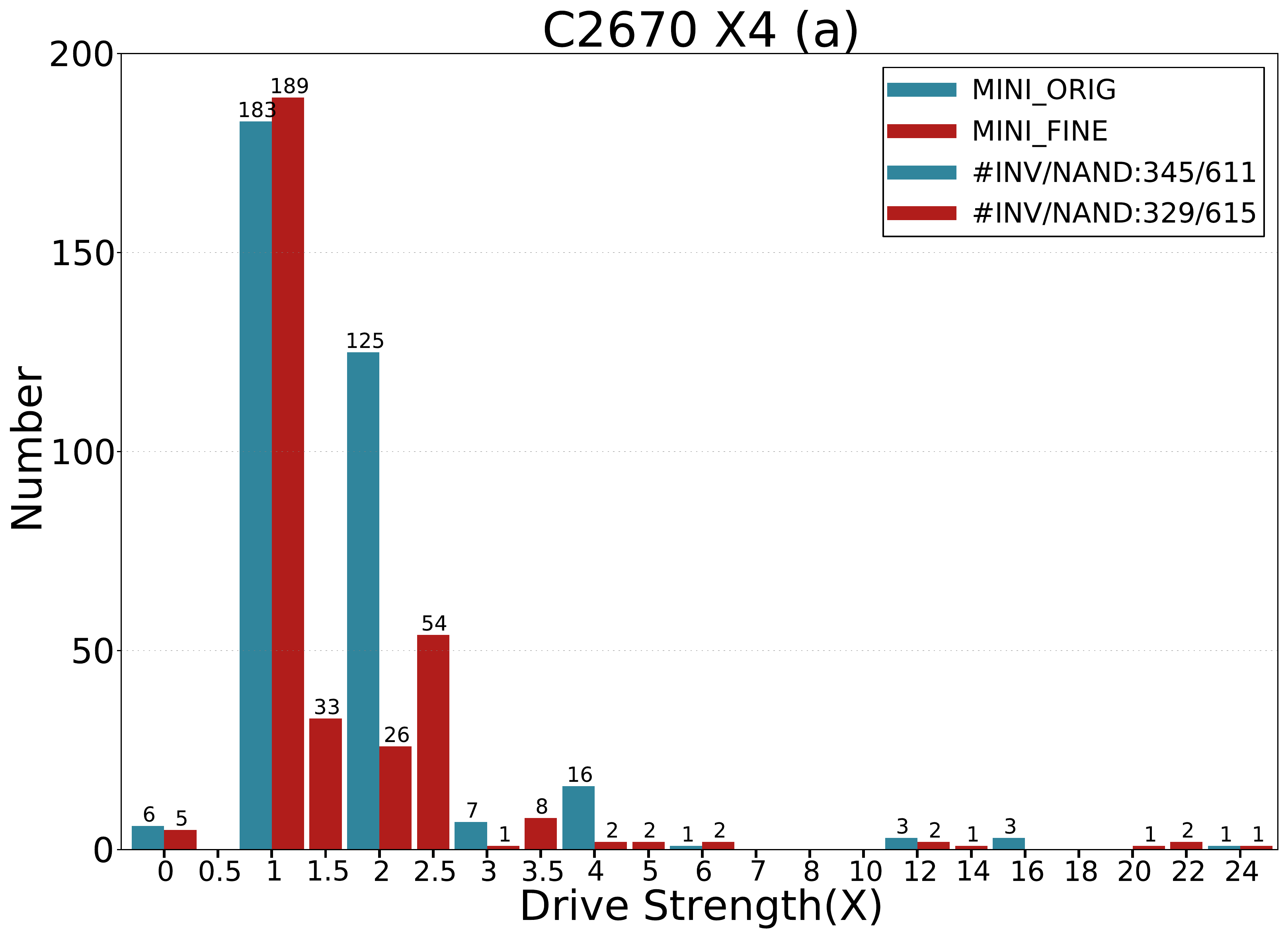}}
     \subfloat{\includegraphics[width=0.325\linewidth]{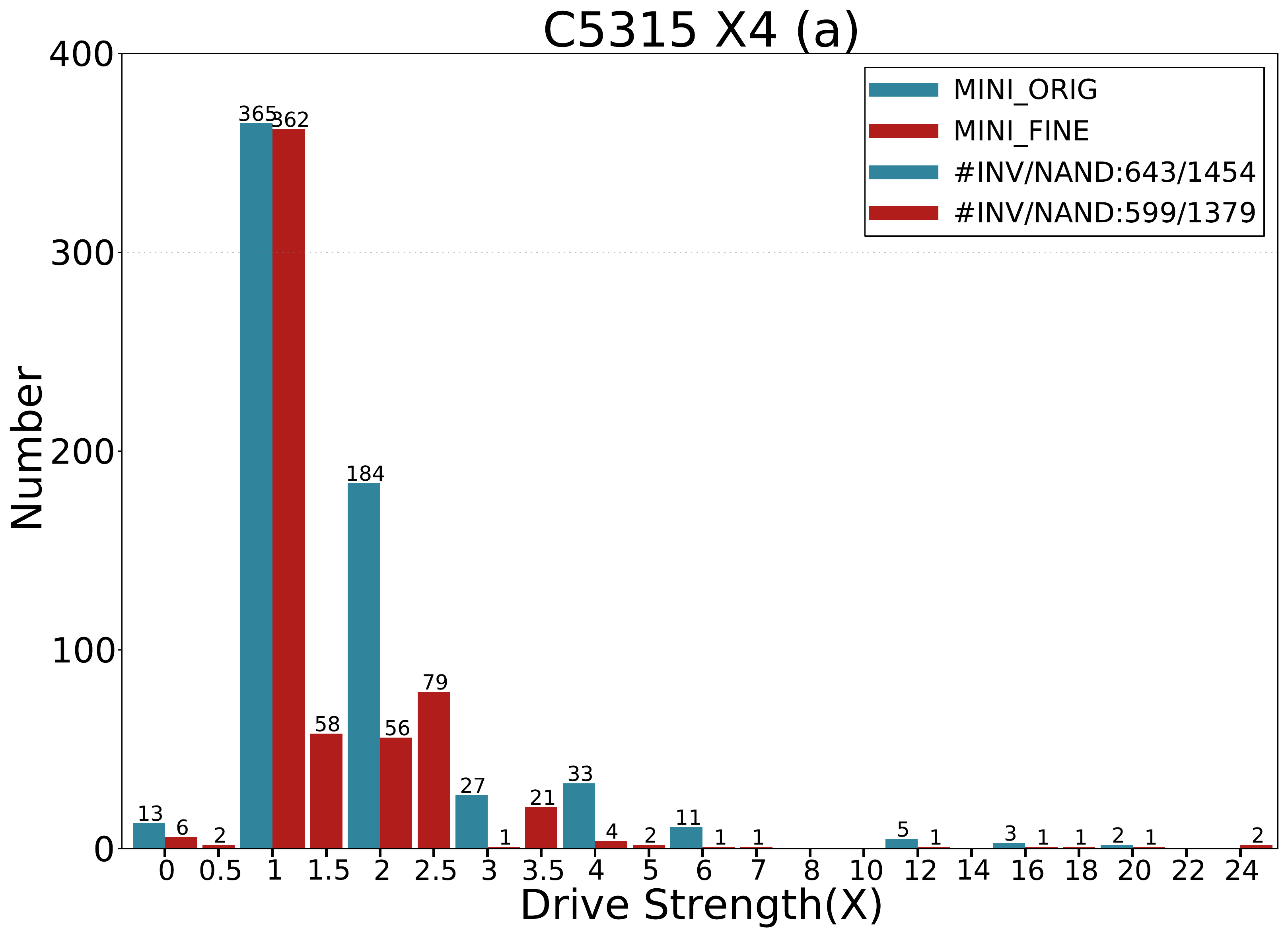}}\\\vspace{-0.3cm}
     \caption{Each plot shows the number of tool-selected inverters' drive strengths of each case. The blue bars are the inverters in original granularity from ``MINI\_ORIG'' and red bars are the fine-grained inverters from ``MINI\_FINE''. The number of synthesised inverters and nand gates are reported in the legends.}
     \label{fig:histogram_results}\vspace{-0.4cm}
\end{figure*}

\subsection{Objective Evaluation in EDA Tools}
The fitness function is set up to simultaneously minimise all objectives ($D_{wc}$, $P_{total}$ and $A_{gate}$), which aims to make a solution perform better in at least one objective without making performance in others worse. All evaluations take place after place-and-route with Innovus\textsuperscript{TM} analysis based on typical corner conditions.

$D_{wc}$: This is the signal propagation time of the critical path, which is equal to the required time minus the worst negative slack amongst all path delays. It is calculated by static timing analysis at the post-route stage.

$P_{total}$: The results from the power analysis in Innovus\textsuperscript{TM}. This is the sum of leakage power, internal power and switching power consumption. The leakage and internal power are summarised in the Liberty (.lib) file as stated in Section~\ref{section:drive strength}-B. Switching power consumed in the charging and discharging of interconnect and load capacitance is calculated based on the equation $\mathrm{P_{switching} = 0.5 \ast C_{L}V^{2}F \ast A}$, where $\mathrm{C_{L}}$ is the output capacitive load (pin capacitance tables are available in Liberty (.lib) file for computing output loading), $\mathrm{V}$ is the supply voltage, $\mathrm{F}$ is frequency, and $\mathrm{A}$ is the average switching activity (the value 0.2 used here is the default from the tool). 

$A_{gate}$: The sum of areas of all logic gates. This is directly reported by the Innovus\textsuperscript{TM}, based on the layout size of the cells used.

\subsection{Multi-threads Running and Runtime}
According to the computing resources and number of licenses available, all experiments in this work are running 24 MOEDA evaluation threads in parallel.

Evolutionary optimisation is population based and requires large numbers of evaluations. In this work, all circuits are placed and routed to achieve accurate metrics as close as possible to sign-off. Such an evaluation needs to be performed for each instance and represents the majority of the overall runtime of the optimisation loop. This can be overcome (sped up) through evaluating instances in the population in parallel using high-performance computing resources. However, each evaluation requires its own license when running solvers in parallel, which makes the achievable degree of parallelism (speedup) dependent on the number of EDA tool licenses available. In addition, the MOEDA flow delivers an entire set of trade-off solutions spanning the feasible design space in one go, rather than just a single, case-specific solution. So the runtime is not the key focus in this work.

\section{Experimental Results}
\label{section:results}

\subsection{Original vs. Fine-grained Cells in Standard Digital Flow}
We firstly load both ``MINI\_ORIG'' and ``MINI\_FINE'' libraries into the standard digital flow to investigate how the tools deal with different drive-granularity libraries and which drive strengths that the tool prefers. Three benchmarks with different circuit structures and functions from ISCAS85 benchmark suite are synthesised and implemented in physical layouts. They are: a 16-bit error detector/corrector (C1908), a 12-bit ALU and controller (C2670) and a 9-bit ALU (C5315). Each circuit is implemented under three different timing constraints and two different output load constraints, resulting in 6 test cases per circuit. This aims to verify that the improved drive-granularity library can demonstrate generic benefits for designs when applying different timing goals (stringent or relaxed) and load capacitance (nominal or larger). The experiment information is summarised in Table~\ref{table:testing_case_summary}.

\begin{table}[htbp]\vspace{-0.2cm}
\begin{center}
\caption{Test Case Summary}\vspace{-0.2cm}
\begin{tabular}{|c|c|c|c|}
  \hline
  Design & Lib & Load & (\#) Required Timing [$ns$]\\
  \hline
  C1908 & ORIG/FINE & X1/X4 & $(a)1.25 \hspace{0.2cm} (b)1.40 \hspace{0.2cm} (c)1.55$ \\
  \hline
  C2670 & ORIG/FINE & X1/X4 & $(a)1.20 \hspace{0.2cm} (b)1.35 \hspace{0.2cm} (c)1.50$ \\
  \hline
  C5315 & ORIG/FINE & X1/X4 &$ (a)1.35 \hspace{0.2cm} (b)1.50 \hspace{0.2cm} (c)1.65$ \\
  \hline
\end{tabular}\vspace{-0.4cm}
\label{table:testing_case_summary}
\end{center}
\end{table}

Fig.~\ref{fig:histogram_results} presents histograms of inverters used in each tool-synthesised benchmark circuit when applying tightest timing requirements from case (a). The blue bars represent the histogram of the ``MINI\_ORIG'' library and the red ones show the histogram of the ``MINI\_FINE'' library. The drive strengths X1 and X2 are the most commonly used cells. They are most dominant in the histograms of all test cases using the ``MINI\_ORIG'' library. A number of fine-grained drive strength inverters are selected by the synthesis tool when using the ``MINI\_FINE'' library. The peak around drive strength X2 is significantly flatter when fine-grained inverters are selected, although the number of drive strength X1 is still high. The likely reason for this is that, for many circuit paths, drive strength X1 is capable of driving the load at the endpoint, which is often a single gate. In addition, the improved drive strength resolution around X1 is exploited, although it may still not be fine enough to reduce the dominant X1 peak in the histograms of inverters used. 

Fig.~\ref{fig:histogram_results} shows that the most-used fine-grained inverters are X1.5, X2.5, X3.5 when ``MINI\_FINE'' library is used. This indicates which fine-grained gate sizes will be most useful and show significant benefits to designs particularly when applying this interpolation method to more common logic functions, e.g., NAND, NOR, AND, etc. Therefore, adding non-integer gate sizes between X0 and X4 (i.e., predominantly-selected by the tool) will be promising for better PPA metrics during synthesis of real-world chip design process, whereas the provided drive options (i.e., normally integer sizes) in foundry libraries are relatively coarse-grained.

All circuit evaluations in terms of PPA are performed based on the physical layouts. Table~\ref{table:iscas} summarises the PPA metrics ($D_{wc}$, $P_{total}$,  $A_{gate}$) for each test case. The normalised ($N.$) results are shown for easier improvement comparison. Each test case has three sets of results that are (1) ``STD+ORIG'': synthesising and implementing designs using the standard flow and the ``MINI\_ORIG'' library; (2) ``STD+FINE'': synthesising and implementing designs using the standard flow with the ``MINI\_FINE'' library; (3) ``MOEDA+FINE'': optimising designs using the MOEDA flow with the ``MINI\_FINE'' library starting from the ``STD+FINE'' results. The results of ``STD+ORIG'' and ``STD+FINE'' are discussed first, followed by an illustration of the ``MOEDA+FINE'' results in the next section.

Based on the results presented in Table~\ref{table:iscas}, using the ``MINI\_FINE'' library can generate designs that achieve better trade-off solutions in PPA compared with using ``MINI\_ORIG'' library running in the standard digital flow, although degradation occurred in one of objectives in some cases, e.g., C1908-X1-(b), C2670-X4-(a) and C5315-X4-(b). This may be due to the richer library leading to a larger design search space and the therefore increased computational complexity increasing synthesis and implementation effort.

It explicitly shows that the synthesis tools can take advantage of the full capabilities of the fine-grained library ``MINI\_FINE'' evidenced by the large amount of fine-grained inverters used, as shown in column ``FINE INV UI.'' (i.e., fine-grained inverters utilisation).

\begin{figure}[ht]
  \centering
     \subfloat{\includegraphics[width=0.7\linewidth]{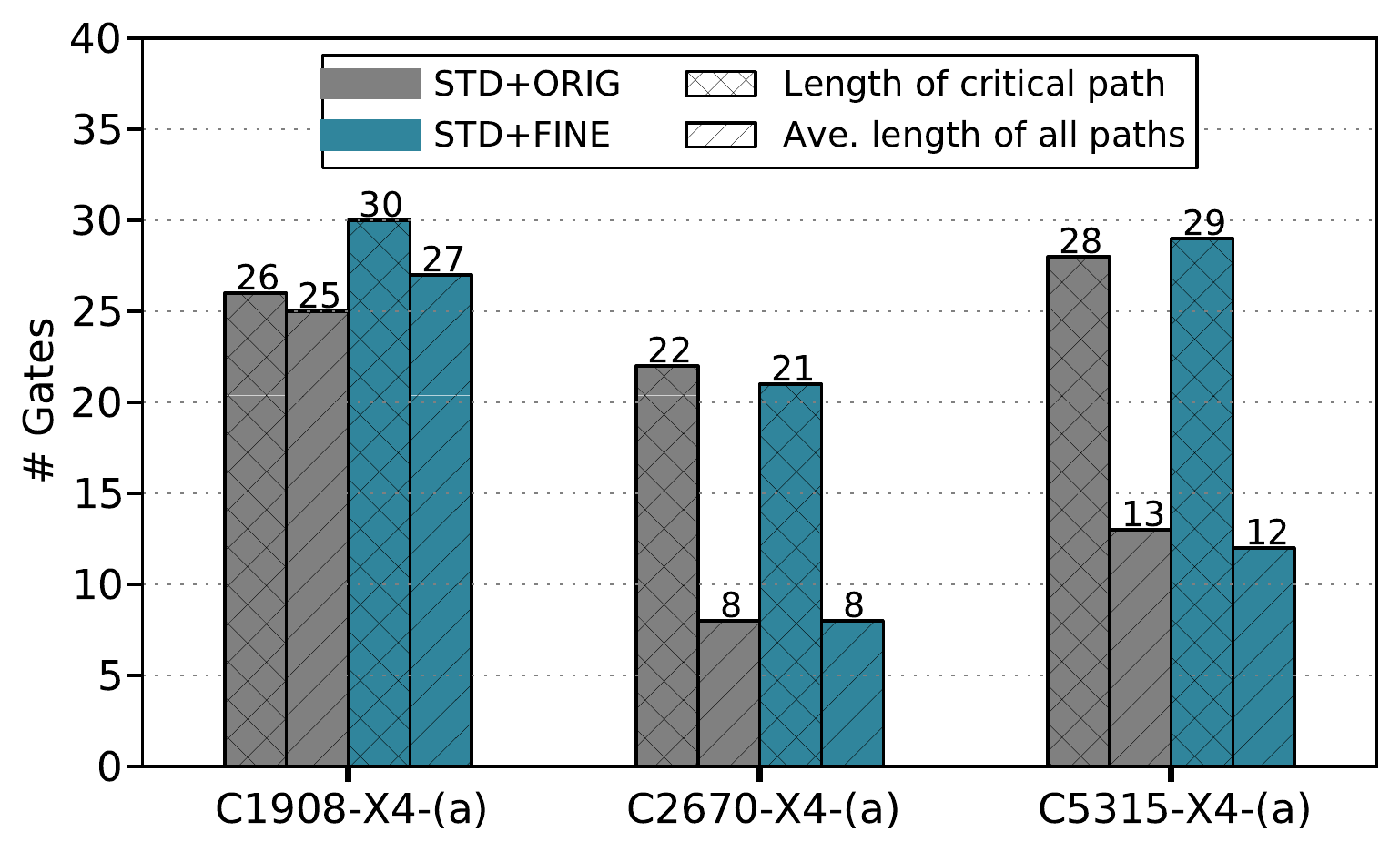}}\vspace{-0.3cm}
     \caption{This shows the changes of circuit paths of each circuit after applying ``MINI\_FINE'' library under the standard flow. The path length is achieved by calculating the gate count of a path.}
     \label{fig:structure_changes}\vspace{-0.4cm}
\end{figure}

The number of inverters, NANDs and total number gates are also reported in Table~\ref{table:iscas} to show how their utilisation changes when synthesising designs using different drive granularity libraries. The total number of gates has decreased in most cases, and up to 6\% (119 gates) reduction in the case of C5315-X4-(a), directly saving circuit area.

Synthesising designs with different granularity libraries may produce solutions with different circuit structures. Fig.~\ref{fig:structure_changes} investigates whether applying fine-grained drive strength cells will change the circuit structure when using the standard flow. So a comparison is made here between the results of ``STD+ORIG'' and ``STD+FINE'' in the X4-(a) case of each circuit. The length of the critical path and average length of all paths are plotted here. Slight difference are shown between ``STD+ORIG'' and ``STD+FINE'' in terms of the circuit paths. This confirms when applying fine-grained drive cells in the standard flow, the synthesis tool chose more suitable cells (fine-grained ones in a significant amount) from a wider available set to meet timing of each path, but the whole circuit structure did not change too much.

\subsection{Fine-grained Cells in MOEDA Flow}
To further improve solutions while balancing multiple objectives, which the standard digital flow is not capable of and instead prioritises timing alone, the MOEDA flow is used to enlarge the solution space, offering a wide range of Pareto-optimised solutions. Subsequent optimisation performed by the MOEDA flow is starting from a set of solutions obtained by the standard digital flow with the ``MINI\_FINE'' library (``STD+FINE''). This is because the results of ``STD+FINE'' achieve better circuit evaluation metrics than the solution of using the ``MINI\_ORIG'' library initially, so that the MOEDA's optimisation efforts are focused on finding better trade-off solutions regarding PPA, rather than starting from scratch.

\begin{figure}[ht]
  \centering\vspace{-0.4cm}
     \subfloat{\includegraphics[width=0.5\linewidth]{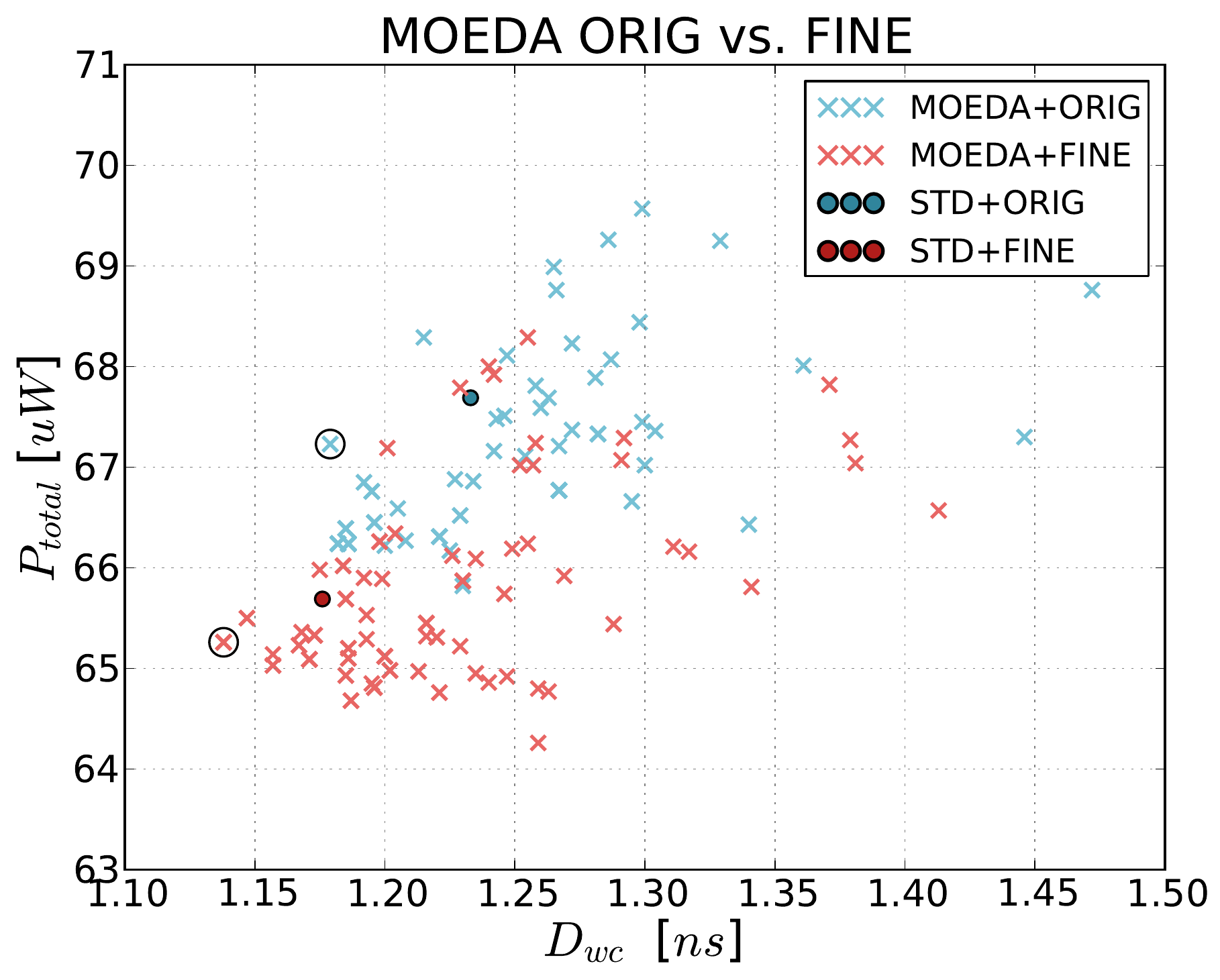}}
     \subfloat{\includegraphics[width=0.5\linewidth]{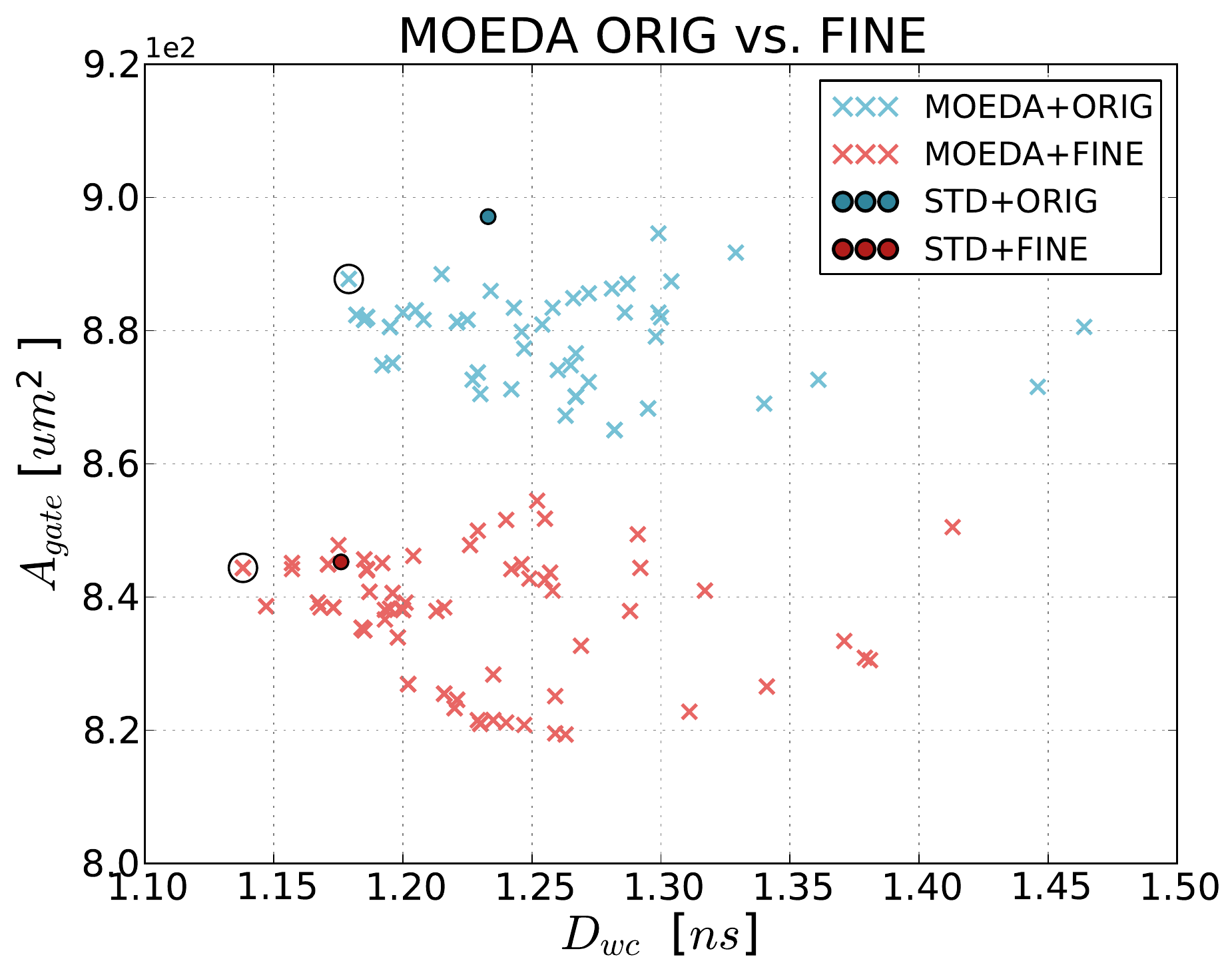}}\vspace{-0.2cm}
     \caption{The MOEDA flow optimisation results comparison between seeding with ``STD+ORIG'' (blue) and ``STD+FINE'' (red). The circled solutions are the best delay solution of each cluster.}\vspace{-0.2cm}
  \label{fig:c1908_X1_a_finevsorig}
\end{figure}

Fig.~\ref{fig:c1908_X1_a_finevsorig} compares the optimisation results of MOEDA flow using the ``MINI\_ORIG'' and the ``MINI\_FINE'' libraries in the C1908-X1-(a) case. Both run with $N$=100 individuals of a population for $M$=100 generations using mutation rate $\rho$=0.5\%. The optimisation run seeded with ``STD+FINE'' solutions can achieve a wider coverage of the design space featuring solutions with better PPA metrics than those based on ``STD+ORIG'' alone. 
\begin{figure}[ht]
  \centering
     \subfloat{\includegraphics[width=0.75\linewidth]{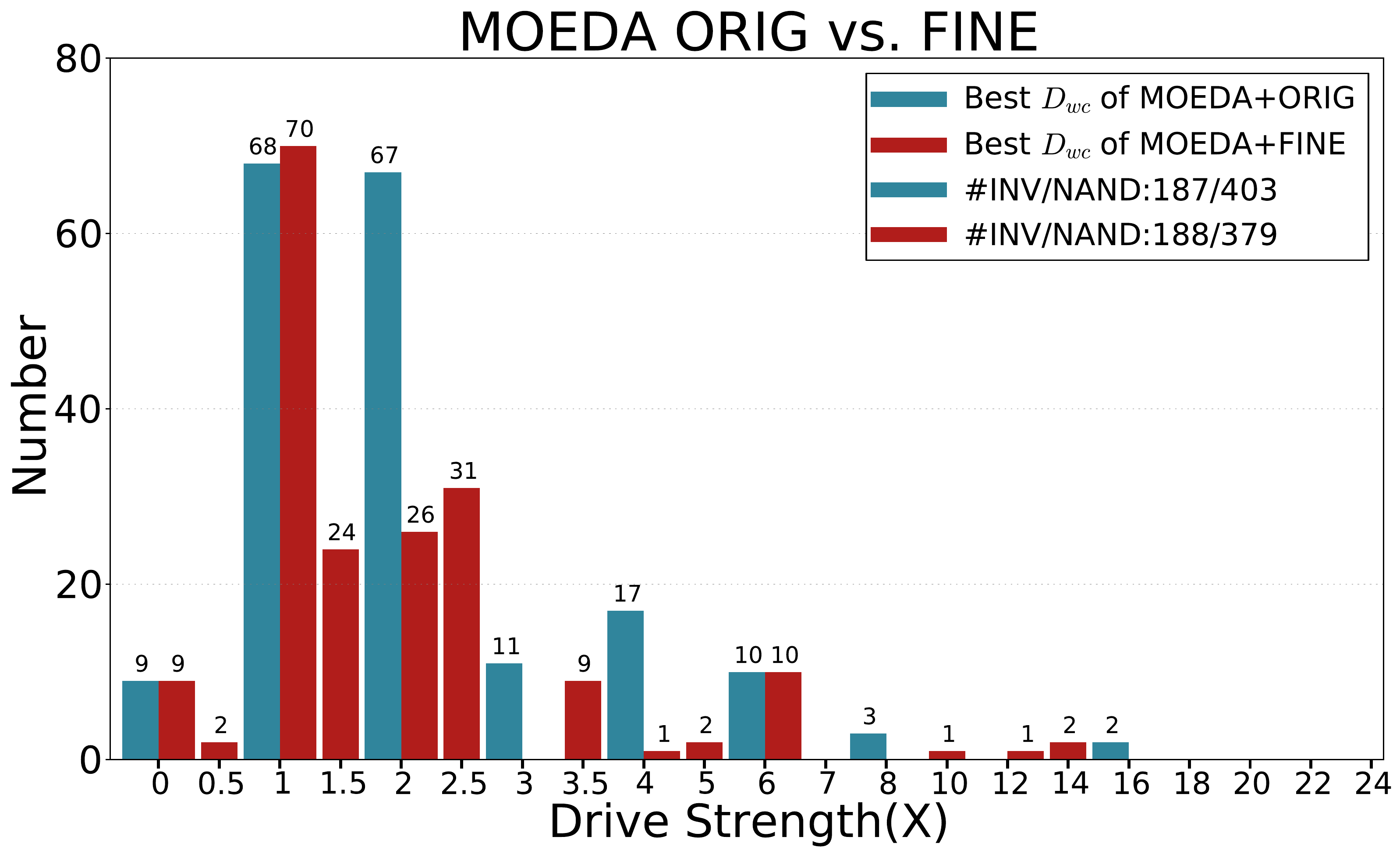}}\vspace{-0.3cm}
     \caption{The inverter histogram of the best delay solution of ``MOEDA+ORIG'' solution space and ``MOEDA+FINE'' solution space from Fig.~\ref{fig:c1908_X1_a_finevsorig}.}\vspace{-0.4cm}
  \label{fig:c1908_X1_a_finevsorig_his}
\end{figure}
Fig.~\ref{fig:c1908_X1_a_finevsorig_his} shows the inverter histogram of each best-delay solution of the ``MOEDA+ORIG'' solution space (in blue bars) and the ``MODEA+FINE'' solution space (in red bars). Both are circled as shown in the plots. This histogram shows similar drive strength distribution compared to the one of C1908-X1-(a) from Fig.~\ref{fig:histogram_results}, which shows the synthesis results without the MOEDA flow. But the drive strength selection results from the tool still has been refined after applying the optimisation of MOEDA flow. In both cases of ``MOEDA+ORIG'' and ``MOEDA+FINE'', drive strengths smaller than X1 are selected more often and drive strengths larger than X2.5 are used less, resulting in power saving. Comparing the solution space of the ``MOEDA+ORIG'' with the ``MOEDA+FINE'' 's,  the ``MOEDA+FINE'' 's results further reduce the use of drive strengths larger than X2.5, so that the power and area of the best $D_{wc}$ solution of ``MOEDA+FINE'' is much lower than the ``MOEDA+ORIG'' 's. This confirms that the MOEDA flow can balance multiple objectives through selection of more appropriate drive strengths for digital circuits. Also, the MOEDA flow can efficiently deal with the richer drive granularity library in trading off solutions.

\begin{table*}[htbp]
\centering\vspace{-0.4cm}
\caption{Results Comparison}\vspace{-0.3cm}
\begin{tabular}{|c|l|c|c|c|c|c|c|c|c|}
    \multicolumn{10}{r}{$N=100$, $M=100$, $\rho=0.5\%$}\\
    \hline
    Design      &\multicolumn{1}{c|}{(\#) Required} &\multirow{2}{*}{Flow} &Lib    &\#        &Total    &FINE INV  &$D_{wc}$ ($N.$) &$P_{total}$ ($N.$) &$A_{gate}$ ($N.$) \\
    (set\_load) &\multicolumn{1}{c|}{Timing}        &                      &(MINI) &INV/NAND  &Gates &UT.[\%]   &[$ns$]          & [$uW$]            &[$um\textsuperscript{2}$] \\
    \hline
    \multirow{9}{*}{C1908 (X1)} &\multirow{3}{*}{$(a)$ 1.25$ns$} & STD   &ORIG &187/403 &590 &0\%    &1.233 (1.00) &67.69 (1.00) &897.12 (1.00) \\
                                &                                & STD   &FINE &188/379 &567 &37.2\% &1.176 (0.95) &65.69 (0.97) &845.28 (0.94) \\
                                &                                & MOEDA &FINE &188/379 &567 &37.8\% &\textbf{1.138 (0.92)} &\textbf{65.26 (0.96)} &\textbf{844.38 (0.94)} \\\cline{2-10}
                                
                                &\multirow{3}{*}{$(b)$ 1.40$ns$} & STD   &ORIG &192/396 &588 &0\%    &1.211 (1.00) &67.59 (1.00) &893.16 (1.00) \\
                                &                                & STD   &FINE &173/362 &535 &37.0\% &1.244 (1.03) &61.65 (0.91) &797.04 (0.89) \\
                                &                                & MOEDA &FINE &173/362 &535 &35.8\% &\textbf{1.186 (0.98)} &\textbf{60.23 (0.89)} &\textbf{790.92 (0.88)} \\\cline{2-10}
                                
                                &\multirow{3}{*}{$(c)$ 1.55$ns$} & STD   &ORIG &189/386 &575 &0\%    &1.218 (1.00) &67.31 (1.00) &881.64 (1.00) \\
                                &                                & STD   &FINE &183/363 &546 &35.5\% &1.212 (0.99) &62.46 (0.93) &819.36 (0.93) \\
                                &                                & MOEDA &FINE &183/363 &546 &34.4\% &\textbf{1.164 (0.95)} &\textbf{60.96 (0.90)} &\textbf{815.76 (0.92)} \\\cline{1-10}
                           
    \multirow{9}{*}{C1908 (X4)} &\multirow{3}{*}{$(a)$ 1.25$ns$*}& STD   &ORIG &200/385 &585 &0\%    &1.306 (1.00) &71.56 (1.00) &889.56 (1.00) \\
                                &                                & STD   &FINE &186/372 &558 &37.1\% &1.200 (0.92) &67.56 (0.94) &852.12 (0.96) \\
                                &                                & MOEDA &FINE &186/372 &558 &37.1\% &\textbf{1.164 (0.89)} &\textbf{66.75 (0.93)} &\textbf{838.80 (0.94)} \\\cline{2-10}
                                
                                &\multirow{3}{*}{$(b)$ 1.40$ns$} & STD   &ORIG &205/421 &626 &0\%    &1.255 (1.00) &78.08 (1.00) &970.56 (1.00) \\
                                &                                & STD   &FINE &183/372 &555 &38.8\% &1.210 (0.96) &67.20 (0.86) &842.76 (0.87) \\
                                &                                & MOEDA &FINE &183/372 &555 &36.1\% &\textbf{1.191 (0.95)} &\textbf{65.99 (0.84)} &\textbf{824.94 (0.85)} \\\cline{2-10}
                                
                                &\multirow{3}{*}{$(c)$ 1.55$ns$} & STD   &ORIG &205/407 &612 &0\%    &1.247 (1.00) &75.05 (1.00) &938.88 (1.00) \\
                                &                                & STD   &FINE &192/387 &579 &39.6\% &1.255 (1.01) &72.31 (0.96) &887.94 (0.95) \\
                                &                                & MOEDA &FINE &192/387 &579 &41.1\% &\textbf{1.180 (0.94)} &\textbf{70.41 (0.94)} &\textbf{870.12 (0.92)} \\
    \hline	
    \multirow{9}{*}{C2670 (X1)} &\multirow{3}{*}{$(a)$ 1.20$ns$} & STD   &ORIG &317/634 &951 &0\%    &0.982 (1.00) &98.23 (1.00) &1377.72 (1.00) \\
                                &                                & STD   &FINE &336/593 &929 &35.7\% &0.952 (0.97) &95.43 (0.97) &1375.56 (0.99) \\
                                &                                & MOEDA &FINE &336/593 &929 &36.6\% &\textbf{0.902 (0.92)} &\textbf{94.75 (0.96)} &\textbf{1368.36 (0.99)} \\\cline{2-10}
                                
                                &\multirow{3}{*}{$(b)$ 1.35$ns$} & STD   &ORIG &360/633 &993 &0\%    &1.013 (1.00) &103.5 (1.00) &1485.00 (1.00) \\
                                &                                & STD   &FINE &359/627 &986 &34.5\% &0.957 (0.94) &103.1 (0.99) &1464.48 (0.99) \\
                                &                                & MOEDA &FINE &359/627 &986 &35.1\% &\textbf{0.895 (0.88)} &\textbf{102.2 (0.98)} &\textbf{1463.76 (0.98)} \\\cline{2-10}
                                
                                &\multirow{3}{*}{$(c)$ 1.50$ns$} & STD   &ORIG &332/601 &933 &0\%    &1.116 (1.00) &98.29 (1.00) &1430.28 (1.00) \\
                                &                                & STD   &FINE &336/601 &937 &28.9\% &1.007 (0.90) &92.85 (0.94) &1355.22 (0.95) \\
                                &                                & MOEDA &FINE &336/601 &937 &29.2\% &\textbf{0.972 (0.87)} &\textbf{92.13 (0.93)} &\textbf{1355.22 (0.95)} \\\cline{1-10}
                           
    \multirow{9}{*}{C2670 (X4)} &\multirow{3}{*}{$(a)$ 1.20$ns$} & STD   &ORIG &345/611 &956 &0\%    &1.041 (1.00) &100.8 (1.00) &1376.64 (1.00) \\
                                &                                & STD   &FINE &329/615 &944 &30.4\% &0.986 (0.95) &101.9 (1.01) &1373.22 (0.99) \\
                                &                                & MOEDA &FINE &329/615 &944 &31.3\% &\textbf{0.921 (0.88)} &\textbf{101.8 (1.009)} &\textbf{1358.10 (0.98)} \\\cline{2-10}
                                
                                &\multirow{3}{*}{$(b)$ 1.35$ns$} & STD   &ORIG &387/662 &1049 &0\%    &1.073 (1.00) &115.4 (1.00) &1559.16 (1.00) \\
                                &                                & STD   &FINE &347/616 &963  &30.5\% &1.046 (0.97) &101.0 (0.88) &1403.28 (0.90) \\
                                &                                & MOEDA &FINE &347/616 &963  &32.0\% &\textbf{0.937 (0.87)} &\textbf{100.0 (0.86)} &\textbf{1402.56 (0.90)} \\\cline{2-10}
                                
                                &\multirow{3}{*}{$(c)$ 1.50$ns$} & STD   &ORIG &338/644 &982 &0\%     &1.083 (1.00) &107.0 (1.00) &1458.72 (1.00) \\
                                &                                & STD   &FINE &331/580 &911 &37.5\%  &1.081 (0.99) &99.07 (0.93) &1400.58 (0.96) \\
                                &                                & MOEDA &FINE &331/580 &911 &38.7\%  &\textbf{0.981 (0.90)} &\textbf{98.64 (0.92)} &\textbf{1379.34 (0.94)} \\
    
    \hline	
    \multirow{9}{*}{C5315 (X1)} &\multirow{3}{*}{$(a)$ 1.35$ns$ *}& STD   &ORIG &641/1442 &2083 &0\%    &1.405 (1.00) &249.4 (1.00) &2968.20 (1.00) \\
                                &                                 & STD   &FINE &607/1399 &2006 &25.0\% &1.417 (1.01) &234.9 (0.94) &2850.66 (0.96) \\
                                &                                 & MOEDA &FINE &607/1399 &2006 &24.9\% &\textbf{1.291 (0.92)} &\textbf{232.0 (0.93)} &\textbf{2841.12 (0.95)} \\\cline{2-10}
                                
                                &\multirow{3}{*}{$(b)$ 1.50$ns$}  & STD   &ORIG &624/1416 &2040 &0\%    &1.433 (1.00) &238.9 (1.00) &2875.32 (1.00) \\
                                &                                 & STD   &FINE &620/1378 &1998 &23.4\% &1.408 (0.98) &234.6 (0.98) &2820.60 (0.98) \\
                                &                                 & MOEDA &FINE &620/1378 &1998 &23.9\% &\textbf{1.332 (0.93)} &\textbf{229.5 (0.96)} &\textbf{2814.48 (0.97)} \\\cline{2-10}
                                
                                &\multirow{3}{*}{$(c)$ 1.65$ns$}  & STD   &ORIG &624/1374 &1998 &0\%    &1.437 (1.00) &235.9 (1.00) &2821.68 (1.00) \\
                                &                                 & STD   &FINE &622/1371 &1993 &22.8\% &1.469 (1.02) &231.5 (0.98) &2801.16 (0.99) \\
                                &                                 & MOEDA &FINE &622/1371 &1993 &24.0\% &\textbf{1.325 (0.92)} &\textbf{226.5 (0.96)} &\textbf{2798.46 (0.99)} \\\cline{1-10}
                           
    \multirow{9}{*}{C5315 (X4)} &\multirow{3}{*}{$(a)$ 1.35$ns$ *}& STD   &ORIG &643/1454 &2097 &0\%    &1.486 (1.00) &259.2 (1.00) &3018.60 (1.00) \\
                                &                                 & STD   &FINE &599/1379 &1978 &27.4\% &1.413 (0.95) &236.5 (0.91) &2812.50 (0.93) \\
                                &                                 & MOEDA &FINE &599/1379 &1978 &27.7\% &\textbf{1.342 (0.90)} &\textbf{236.4 (0.91)} &\textbf{2808.36 (0.93)} \\\cline{2-10}
                                
                                &\multirow{3}{*}{$(b)$ 1.50$ns$ *}& STD   &ORIG &605/1357 &1962 &0\%    &1.592 (1.00) &244.3 (1.00) &2784.24 (1.00)\\
                                &                                 & STD   &FINE &604/1372 &1976 &25.7\% &1.543 (0.97) &237.6 (0.97) &2798.28 (1.005)\\
                                &                                 & MOEDA &FINE &604/1372 &1976 &25.5\% &\textbf{1.356 (0.85)} &\textbf{231.6 (0.95)} &\textbf{2796.30 (1.004)} \\\cline{2-10}
                                
                                &\multirow{3}{*}{$(c)$ 1.65$ns$}  & STD   &ORIG &652/1457 &2109 &0\%    &1.343 (1.00) &255.9 (1.00) &2989.08 (1.00) \\
                                &                                 & STD   &FINE &619/1397 &2016 &24.1\% &1.365 (1.02) &242.2 (0.95) &2870.82 (0.96) \\
                                &                                 & MOEDA &FINE &619/1397 &2016 &24.7\% &\textbf{1.268 (0.94)} &\textbf{239.7 (0.93)} &\textbf{2863.26 (0.95)} \\
    \hline
\end{tabular}\vspace{-0.4cm}
\label{table:iscas}
\end{table*}

Due to the previous findings, MOEDA flow optimisation is carried out for the next experiments, only initialised with ``STD+FINE'' seed solutions from the standard flow. All the rest of experiments also run with 100 individuals for 100 generations using 0.5\% mutation rate. The MOEDA optimisation results highlighted in Table~\ref{table:iscas} are the best trade-off solutions from the entire final solution space. The best trade-off solution is defined here as an individual from the final generation that is positioned at the shortest Euclidean distance from the origin. These trade-off solutions demonstrate the optimisation capability of achieving improvements in all objectives simultaneously. Four test cases marked with stars represent that ``STD+ORIG'' solutions have already failed timing requirements. Most of these failed cases have been improved with better $D_{wc}$ in ``STD+FINE'' solutions, and all of them have been optimised by the MOEDA flow without compromising on other objectives to the point that they achieve timing closure.

\begin{figure}[ht] 
  \centering
     \subfloat{\includegraphics[width=0.9\linewidth]{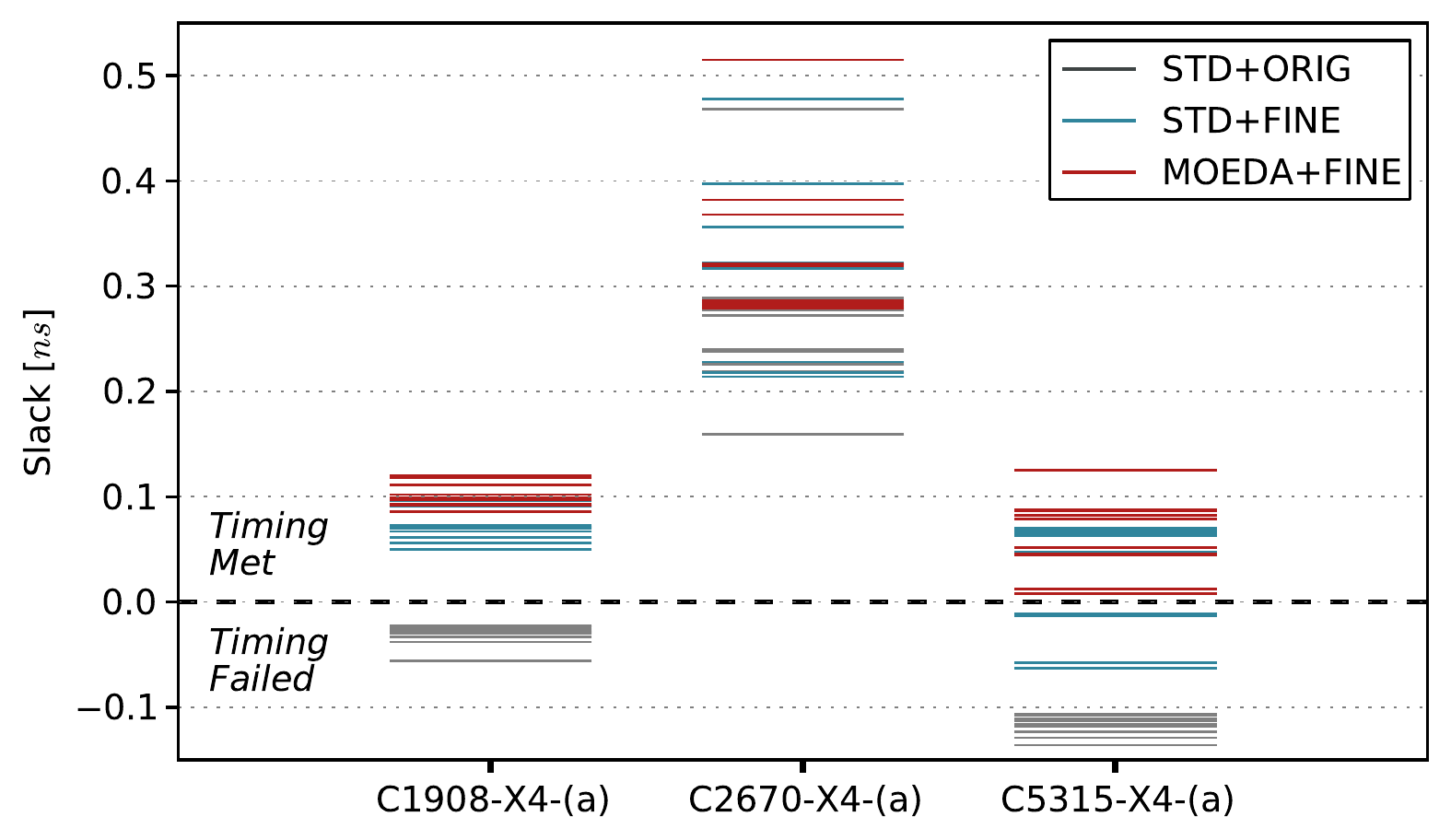}}\vspace{-0.3cm}
     \caption{Ten worst timing paths of test circuits for each corresponding tight timing constraint case X4-(a). All paths above the dash line have positive slacks which meet the timing.}
     \label{fig:timing_path_analysis}\vspace{-0.5cm}
\end{figure}

Fig.~\ref{fig:timing_path_analysis} plots the ten worst timing paths of the X4-(a) case (tightest timing constraint in this work) of each test circuit. This shows how the timing of paths, particularly the critical path, has been optimised. The results of ``MOEDA+FINE'' (red lines) recovered the all timing failed paths and performed the hill-climbing on the slack of critical paths, where only applying ``MINI\_FINE'' library in the standard (STD) flow (``STD+FINE'' in blue lines) is not capable of. In addition, the results of applying ``MINI\_ORIG'' in the STD flow (``STD+ORIG'' in gray lines) explicitly show inferior timing performance, especially in C1908 and C5315 circuits with negative slacks.

\begin{figure}[ht] 
   \centering\vspace{-0.3cm}
      \subfloat{\includegraphics[width=0.85\linewidth]{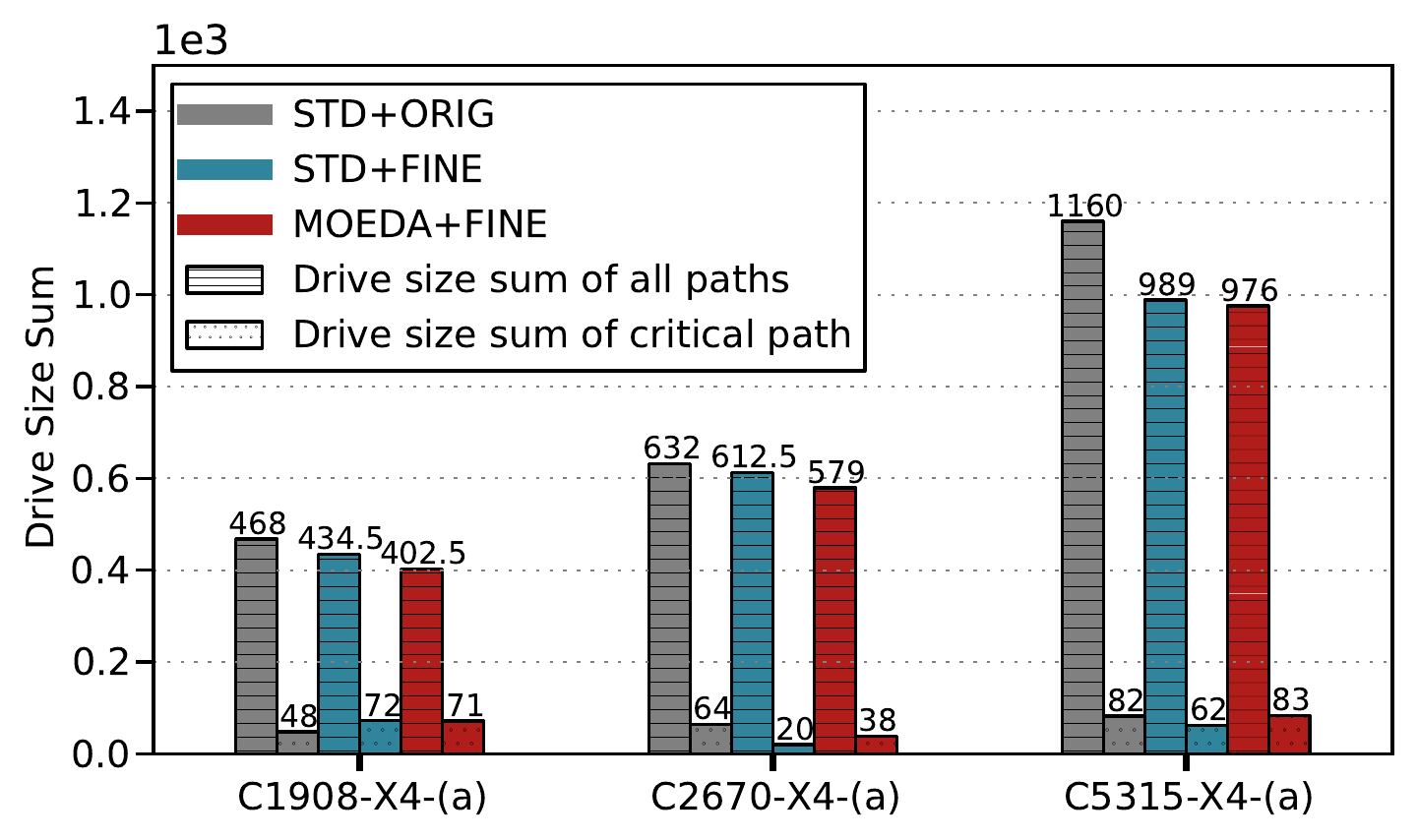}}\vspace{-0.3cm}
      \caption{The changes of drive strengths of critical paths and overall circuits when applying ``STD+ORIG'', ``STD+FINE'' and ``MOEDA+FINE''. The sum of drive strengths are reported from the X4-(a) case of each benchmark.}
      \label{fig:path_size}
 \end{figure}

To investigate the changes of drive strength selection when using different libraries and flows, Fig.~\ref{fig:path_size} presents the sum of drive strength sizes of each whole circuit and their corresponding critical paths. The tight timing case X4-(a) of each benchmark is still used for analysis here. Based on the observation of this plot, the overall drive size sum of all circuit paths has decreased after applying ``STD+FINE'' and has further been optimised by the MOEDA flow. This straightforwardly saves the resulting power and area of designs.

In terms of critical paths relating to the circuit worst slack, more larger drive cells from ``MINI\_FINE'' library are selected by MOEDA flow to solve timing violations in C1908 and C5315, since they had timing failed paths in the initial solution generated by ``STD+ORIG''. The critical path delay of ``STD+FINE'' solution in C1908 and C5315 was improved over ``STD+ORIG'', but the total size of selected drive strengths is not always increased. This indicates that for each timing path drive strengths need to be optimised rather than simply scaling up the gate sizes. In C2670 circuit, the drive strength sum of the critical path in ``STD+FINE'' solution is greatly reduced while all paths are meeting the timing constraint and the worst slack is getting improved. The MOEDA flow then selects more larger cells to push the timing performance but the used drive strengths is still less than the ``STD+ORIG'' one.

In addition, since margins shown in the EDA tools for drive strength mapping, that redundant larger cells are selected by the standard flow with using the coarse-grained library, the performance of EDA tools is variable and might be not capable of getting an optimum solution, particularly when handling enlarged search space.

\begin{figure}[ht]
  \centering\vspace{-0.3cm}
     \subfloat{\includegraphics[width=0.5\linewidth]{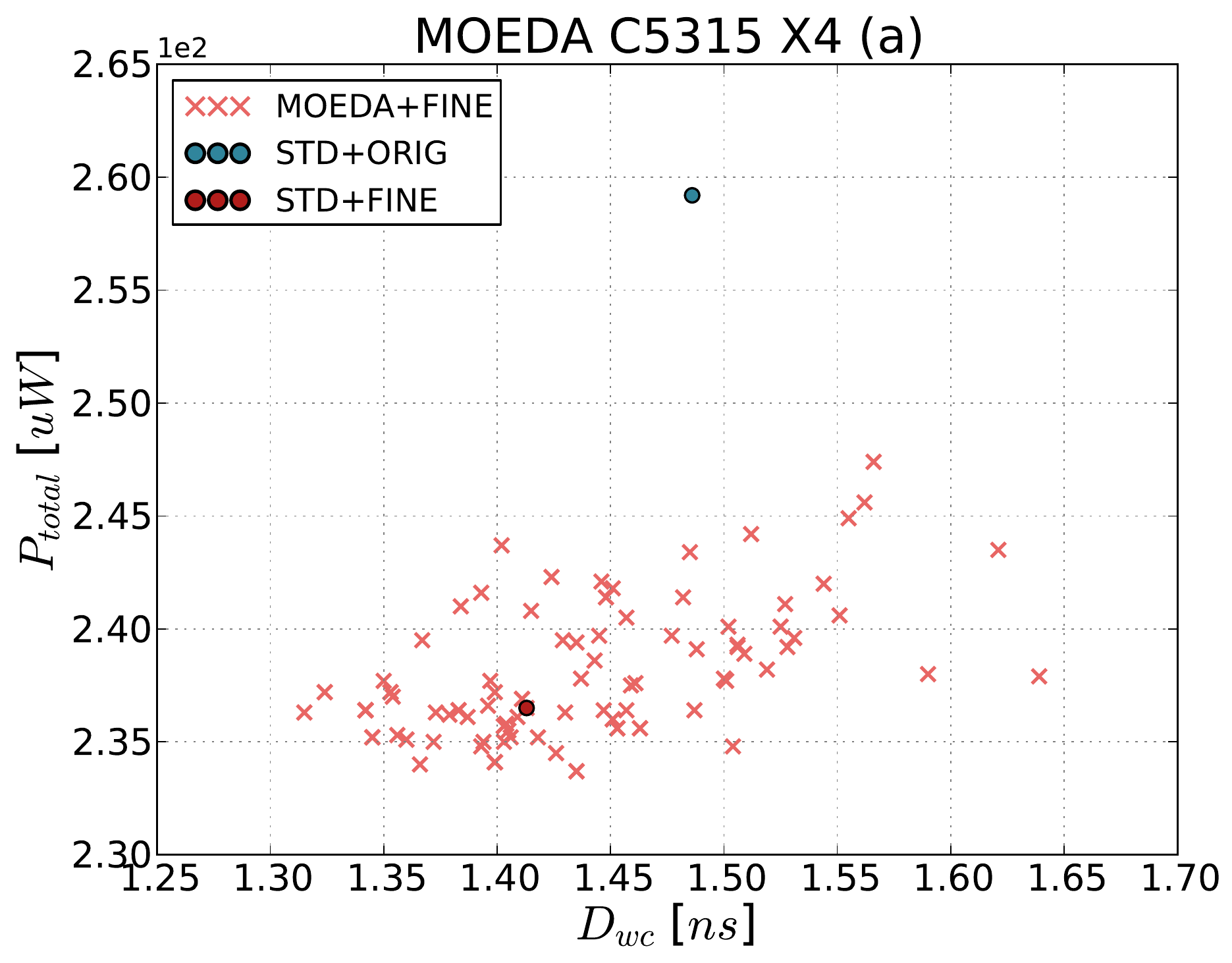}}
     \subfloat{\includegraphics[width=0.5\linewidth]{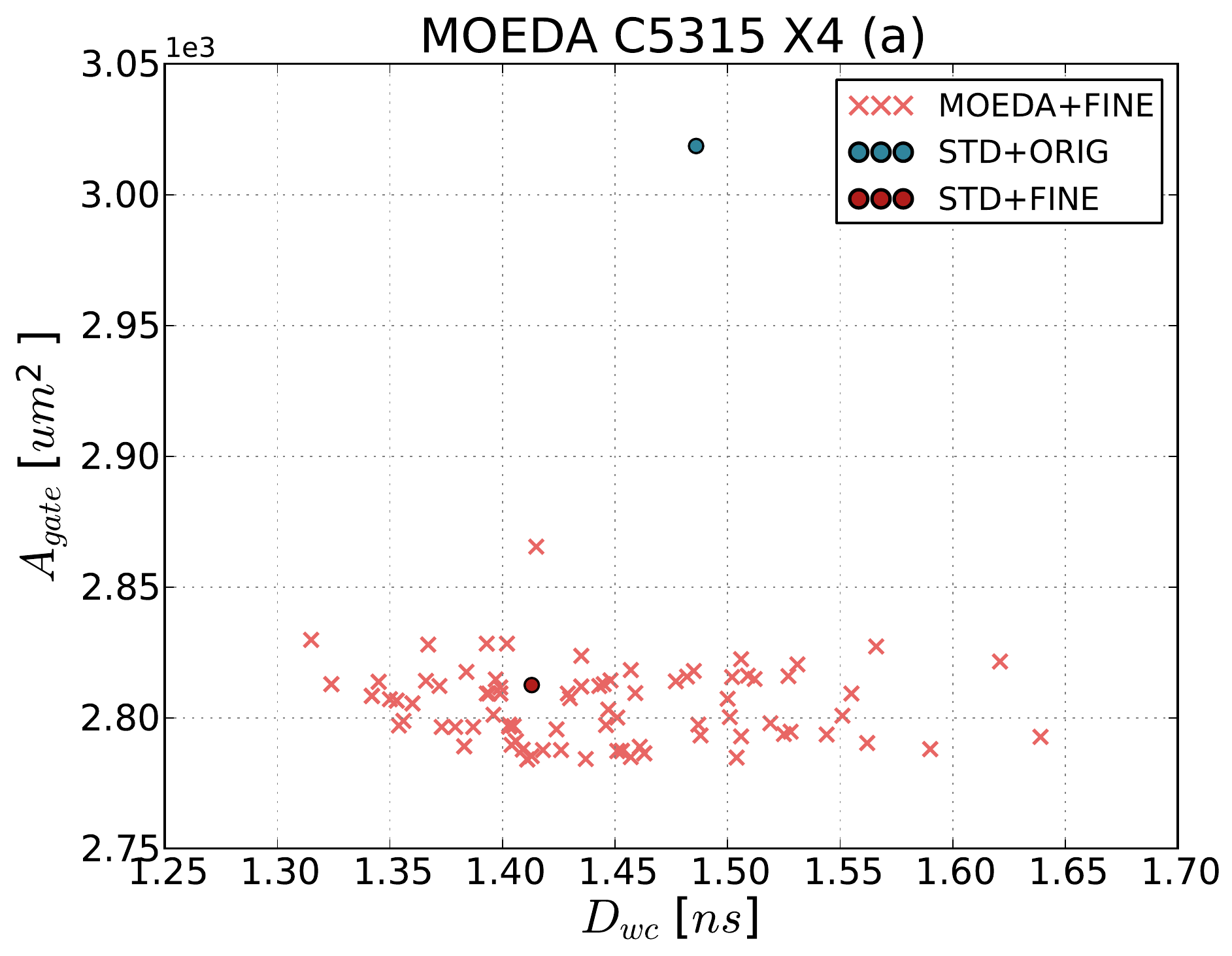}}\\\vspace{-0.2cm}
     \caption{The left plots in ``$D_{wc}$ vs. $P_{total}$'' and ``$D_{wc}$ vs. $A_{gate}$'' is on the right. There are two individuals in the round shape are the ``STD+ORIG'' and ``STD+FINE'' solutions. All other individuals in the shape of cross are the final generation of MOEDA optimised results based on the ``STD+FINE'' solution (MOEA seed).}\vspace{-0.3cm}
  \label{fig:moeda_fine}
\end{figure}

To show all trade-off solutions and optimised solution space, Fig.~\ref{fig:moeda_fine} presents the final generation of MOEDA results of the tight timing constraint case X4-(a) of
the C5315 benchmark. The ``STD+ORIG'' and ``STD+FINE'' solutions are also plotted here comparison. The MOEDA flow has successfully enlarged the feasible solution space while simultaneously achieve significant improvements in all PPA metrics. If designers focus on one or two of these objectives, the available solutions from MODEA flow can obtain greater objective improvements than the trade-off solutions' reported in the Table~\ref{table:iscas}. The runtime of the largest and most complex case C5315-X4-(a) is 5.5 hours.

\section{Conclusion and Future work}
\label{section:conclusion}

We have shown that digital synthesis and implementation tools produced solutions can be improved when provided with fine-grained drive strength cell libraries. The industrial flow exploited the finer drive strengths to improve PPA metrics of all benchmarks used. The results indicate that providing finer drive resolution around predominantly-selected drive strengths (typically between X0 and X4) is particularly useful. This suggests that enriching drive options of functions of a standard cell library around predominantly selected drive strengths is a promising method to get better performance for large-scale designs out of the standard EDA tools.  

The main challenge of the proposed fine-grained cells approach is that enlarged standard cell libraries result in a larger design search space. Hence, EDA tools need to make a greater effort during drive strength mapping, due to the increased computational complexity, and may not always arrive at an optimum solution in a given time frame. The proposed MOEDA digital design flow can overcome these issues as it is capable of further balancing PPA metrics and provide a range of design solutions where standard EDA tool performance is quite variable and cannot trade-off PPA well.

The capability of the proposed MOEDA flow to offer a set of well-balanced, and often improved, trade-off solutions with regard to PPA also opens up opportunities for designers to choose the most appropriate solution for different applications.

Based on these observations, we will expand fine drive strength granularity to more logic functions and investigate whether further performance, power, area (PPA) benefits can be achieved when handling very large circuits or when fitting designs into constrained floor plans and pin layouts. In addition, exploring how different drive strength combinations will affect the PPA metrics, and how to determine an optimum cell sets for possible best design results, will be investigated.

\end{document}